\documentclass[useAMS,usenatbib]{mn2e}

\usepackage{subfigure}
\usepackage{graphicx}
\usepackage{times}
\usepackage{tabularx}
\usepackage{natbib}
\usepackage{url} 
\usepackage{xcolor}
\usepackage{caption}
\usepackage{amsmath}
\usepackage{float}
\usepackage{multirow}

\newcommand{\enzo}{{\it {\small ENZO}}}
\newcommand{\CRaTer}{{\it {\small CRaTer}}}
\newcommand{\dd}{\mathrm{d}}
\newcommand{\Mpc}{\mathrm{Mpc}}
\newcommand{\Msun}{\mathrm{M}_{\odot}}
\newcommand{\Mth}{\mathrm{M}_{200}}

\newcommand{\kpc}{\mathrm{kpc}}

\newcommand{\km}{\mathrm{km}}
\newcommand{\sek}{\mathrm{s}}

\newcommand{\G}{\mathrm{G}}
\newcommand{\K}{\mathrm{K}}
\newcommand{\Gyr}{\mathrm{Gyr}}
\newcommand{\Myr}{\mathrm{Myr}}

\newcommand{\fc}{F_{\mathrm{comp}}}
\newcommand{\fst}{F_{\mathrm{stretch}}}
\newcommand{\fb}{F_{\mathrm{baro}}}
\newcommand{\fdiss}{F_{\mathrm{diss}}}
\newcommand{\fadv}{F_{\mathrm{adv}}}

\begin{document}
 \definecolor{myred}{rgb}{1,0,0}
 
 \title[Vorticity and enstrophy in the intracluster medium]{Evolution of vorticity and enstrophy in the intracluster medium}
 \author[D. Wittor, T. Jones, F. Vazza, M. Br\"{u}ggen]{D. Wittor$^{1}$\thanks{%
 E-mail: dwittor@hs.uni-hamburg.de}, T. Jones$^{2}$, F. Vazza$^{1,3}$, M. Br\"{u}ggen$^{1}$\\
 $^{1}$ Hamburger Sternwarte, Gojenbergsweg 112, 21029 Hamburg, Germany \\
 $^{2}$ University of Minnesota Twin Cities Minneapolis, MN, USA \\
 $^{3}$ INAF, Istituto di Radioastronomia di Bologna, via Gobetti 101, I-41029 Bologna, Italy}
 \date{Accepted ???. Received ???; in original form ???}
 \maketitle

 \begin{abstract}
Turbulence generated by large-scale motions during structure formation affects the evolution of the thermal and non-thermal components of the intracluster medium. \\
As enstrophy is a measure of the magnitude of vorticity, we study the generation and evolution of turbulence by analysing the Lagrangian history of enstrophy. For this purpose we combine cosmological simulations carried out with the \enzo-code with our Lagrangian post-processing tool \CRaTer. This way we are able to quantify the individual source terms of enstrophy in the course of the accretion of groups onto galaxy clusters. Here we focus on the redshift range from $z=1$ to $z=0$. Finally, we measure the rate of dissipation of turbulence and estimate the resulting amplification of intracluster magnetic fields.\\
We find that compressive and baroclinic motions are the main sources of enstrophy, while stretching motions and dissipation affect most of the ensuing enstrophy evolution. The rate of turbulent dissipation is able to sustain the amplification of intracluster magnetic fields to observed levels.
 \end{abstract}
 \label{firstpage}
 \section{Introduction}\label{sec:intro}
 The intracluster medium (ICM) is a hot ($T \sim 10^7-10^8$ K), dilute plasma that hosts turbulent motions across all scales. Turbulence is driven on cluster scales, $\sim$ few $\Mpc$, as gravitational energy is converted into kinetic energy during the process of hierarchical structure formation \citep[see][and references therein for a recent review]{2015ASSL..407..599B}. Accretion flows convert their kinetic energy into turbulent motions through tangential flows, fluid instabilities or baroclinic motions. The turbulence then cascades from driving scales to dissipative scales and heats the plasma, (re-)accelerates cosmic-ray particles and amplifies magnetic field \citep[e.g.][]{2007MNRAS.378..245B,2015Natur.523...59M, 2015A&C.....9...49S}. Turbulence can also be driven on galactic scales, $\sim 10 \ \kpc$, for example by outflows driven by active galactic nuclei (AGN) or ICM-based magneto-thermal instabilities \citep[e.g.][]{2012ApJ...750..166M,2013ApJ...762...69Z}.\\
In this work, we are tracking the turbulence associated with substructures that are accreted by clusters at $z < 1$. These are typically groups with typical masses of $\sim 10^{13} \ \Msun$, and they are expected to contribute up to $\sim 70 \%$ to the total mass of massive galaxy clusters \citep[e.g.][]{2009ApJ...690.1292B}. \\
 Current observations measure turbulence through the SZ-effect or pressure fluctuations and line spectroscopy in X-ray  \citep[e.g.][]{2016MNRAS.463..655K,2015A&A...575A..38P,2016MNRAS.458.2902Z}. Future X-ray observations should be able to detect the driving scale of turbulence directly due to the outstanding spectral resolution of the new generation of telescopes (e.g. \texttt{Athena}). 
 The analysis of the turbulent motions is rendered difficult by the need to isolate uncorrelated flows from, both correlated flows on large scales ($\ge 0.1 - 1 \ \Mpc$) and small-scale velocity perturbations produced by shocks.  Turbulence is also dependent on the local gas conditions, as the compressive turbulent energy can make up only a few percent or up to $15-30$ per cent of the total turbulent kinetic energy. This  is important for example, for the understanding of cosmic-ray acceleration. The compressive turbulent component, e.g. curl-free component, most likely follows a Burgers-like spectrum, which reduces the power for cosmic-ray acceleration\footnote{In the case the magnetosonic waves, that are responsible for the acceleration of particles, are dissipated at shocks steepening the cascade and reducing the effective energy transfer to the particles \citep[][]{2015ApJ...800...60M}.} \citep[][]{Brunetti_Jones_2014_CR_in_GC,2015ApJ...800...60M}. \\
 \citet{2015ApJ...810...93P} simulated the properties of MHD turbulence driven by various combinations of solenoidal and compressive processes. Their objective was to understand the physical sources of ICM enstrophy (see Sec. \ref{ssec:enstrophy}) and the associated turbulent amplification of magnetic fields. \cite{2017MNRAS.464...210V} extended this work by analysing a major merger cluster, finding that enstrophy is generated by baroclinic and shock-related motions during accretion and merger processes. In the cluster interior, vortex stretching seeded by mergers is enhancing and generating enstrophy. \\
 In this work, we use our post-processing tool \CRaTer \ to analyse the Lagrangian evolution of enstrophy in eight different clusters taken from the Itasca Simulated Clusters (ISC).  The paper is structured as follows: After summarizing the most important points of enstrophy generation and evolution in Sec.~\ref{ssec:enstrophy}, we will give detailed information on our simulations and numerical tools in Sec.~\ref{sec:setup}. In Sec.~\ref{ssec:general}, we give an overview of the general properties of our cluster sample. Our results on the Lagrangian evolution of enstrophy are presented in \ref{ssec:it90_3_lagrangian}. We focus on the evolution of growth and decay times associated with the different source terms that generate enstrophy in Sec.~\ref{ssec:it90_3_tturn}. In Sec.~\ref{ssec:diss}, we give estimates on the turbulent energy dissipation and the corresponding magnetic field amplification. Finally, we summarise our results and conclude in Sec.~\ref{sec:summary}. In the Appendix, we further give an analytical derivation of how the dissipation rate of turbulence can be estimated in our simulation.
 \subsection{Evolution of enstrophy}\label{ssec:enstrophy}
 The kinetic energy of turbulence in the ICM is mostly \citep[$60-90$ per cent][]{2015Natur.523...59M} of solenoidal nature (divergence-free) and its amount can be measured by the vorticity $\mathbf{\omega} = \nabla \times \mathbf{v}$. However, the average vector vorticity tends to zero and other proxies for solenoidal turbulence are needed. The enstrophy $\epsilon = \frac{1}{2} \left( \nabla \times \mathbf{v} \right)^2$ is such a proxy as it measures the magnitude of vorticity. The equation for the evolution of enstrophy is derived by taking the dot-product of the vorticity and the vorticity equation \citep[for more details see][]{2015ApJ...810...93P}. The evolution of enstrophy in a fixed, Eulerian frame is determined by \textit{advective}, \textit{compressive}, \textit{stretching} and \textit{baroclinic} motions\footnote{Notice, \citet{2015ApJ...810...93P} include a \textit{magnetic} term in their equation. This term is neglected here as our simulations only use pure hydrodynamics.} as well as \textit{dissipation}:
 \begin{align}
  \left(\frac{\mathrm{d} \epsilon}{\mathrm{d}t}\right)_{\mathrm{euler}} = F_{\mathrm{adv}} + F_{\mathrm{comp}} + F_{\mathrm{stretch}} +  F_{\mathrm{baro}} + F_{\mathrm{diss}} \label{eq:enst_euler_deriv}.
 \end{align}
 The individual sink and source terms (from here on we will refer to them as source terms) are:
 \begin{align}
  F_{\mathrm{adv}} &= -\nabla \cdot (\mathbf{v} \epsilon) = - (\epsilon \nabla \cdot \mathbf{v} + \mathbf{v} \cdot \nabla \epsilon), \label{eq:Fadv}\\
  F_{\mathrm{comp}} &= - \epsilon \nabla \cdot \mathbf{v}  \label{eq:Fcomp},\\
  F_{\mathrm{stretch}} &= 2\epsilon (\hat{\omega} \cdot \nabla ) \mathbf{v} \cdot \hat{\omega},  \label{eq:Fstretch}\\
  F_{\mathrm{baro}} &= \frac{\vec\omega}{\rho^2} \cdot ( \nabla \rho \times \nabla P)  \label{eq:Fbaro},\\
  F_{\mathrm{diss}} &= \nu \vec\omega \cdot \left(\nabla^2 \vec\omega + \nabla \times \mathbf{G}\right), \label{eq:Fdiss} \\
  &\mathrm{with} \ \ \vec\omega  = \nabla \times \mathbf{v} . \label{eq:curl}
 \end{align}
 In the equations above, $\rho$ and $P$ are the gas density and pressure, $\nu$ is the kinematic viscosity and $\mathbf{G} = (1/\rho) \nabla \rho \cdot \mathbf{S}$, with the traceless strain tensor $\mathbf{S}$\footnote{$S_{ij} = (1/2)(u_{ij}+u_{ji})-(1/3)\delta_{ij} \nabla \times \mathbf{u}$} \citep[][]{2006MNRAS.370..415M}. A hat denotes a unit vector. We notice that all derivatives are computed using a second-order central difference.\\
 Each source term represents a different physical process leading to the generation, amplification and destruction of enstrophy. The \textit{advective}, $\fadv$, source term describes conservative advection of enstrophy across the cluster. The \textit{compressive}, $\fc$, source term accounts for both reversible compression and rarefractions as well as enstrophy enhancements due to shock compression. The net influence of shock compression on enstrophy is amplification, although as discussed in \cite[][]{2015ApJ...810...93P} creation of enstrophy within shocks really comes from the strain term in Eq. \ref{eq:Fdiss}, combined with subsequent compression within the shock. The \textit{stretching} source term, $\fst$, accounts for the generation of enstrophy by vortex stretching. \textit{Baroclinic}, $\fb$, generation of enstrophy takes place in baroclinic flows, in which the pressure is not a function of density alone; that is, the flow is not barotropic. In our case, where the gas equation of state is adiabatic, that corresponds to flow with non-uniform entropy, which develops behind complex or unsteady shock structures during cluster formation. The \textit{dissipation}, $F_{\mathrm{diss}}$, term accounts for viscous dissipation of solenoidal flow. The dominant component of the dissipation term corresponds to the damping of turbulent eddies, although the second component can also act as a source term in shocks. For the moment, we ignore $\fdiss$ since we have no explicit viscosity, $\nu$, in our simulations. In Sec. \ref{ssec:diss} and in Appendix \ref{app:math} we will estimate the effective viscosity by looking at the dissipation of solenoidal turbulent energy. Clearly, the baroclinic and the dissipation term (through its strain tensor contribution) are the only source terms that are able to generate vorticity. The other source terms depend on the enstrophy itself and therefore they cannot generate enstrophy from zero. \\
 Eq. \ref{eq:enst_euler_deriv} describes the Eulerian evolution of enstrophy. For the tracer analysis we need to transform this into a Lagrangian frame, moving with the ICM fluid, as the change of enstrophy recorded by the tracers between two consecutive timesteps corresponds to the Lagrangian time derivative of the enstrophy. It is computed from Eq. \ref{eq:enst_euler_deriv} by  adding $\mathbf{v} \cdot  \nabla \epsilon$ and neglecting $\fdiss$, giving
 \begin{align}
  \left(\frac{\mathrm{d} \epsilon}{\mathrm{d}t}\right)_{\mathrm{lagrange}} = 2 \cdot F_{\mathrm{comp}} + F_{\mathrm{stretch}} +  F_{\mathrm{baro}} . \label{eq:enst_lagrange_deriv}
 \end{align}
 Enstrophy has the dimensions of inverse time squared, so is intuitively best understood in terms of characteristic ``turnover rate, or, alternatively, eddy turn over time'' for the turbulence. Similarly, the measurements of each source term in Eq. \ref{eq:Fadv}-\ref{eq:Fdiss} are most simply understood in terms of the turnover time.
 As a measurement for the impact of each source term we compute the local, effective and individual source growth/decay times as \\
\begin{align}
   t_{\mathrm{eff}}(t) &= \frac{\epsilon(t)}{\Sigma_i F_i(t)} \label{eq:Tturn} \\
    t_i &= \frac{\epsilon(t)}{F_i(t)} \label{eq:Tsource}.
\end{align}
 In the equations above the index $i$ refers to the individual source terms.
 \section{Simulation setup}\label{sec:setup}
 \subsection{ENZO}\label{ssec:enzo}
 In this work we study eight galaxy clusters taken from the ISC sample\footnote{http://cosmosimfrazza.myfreesites.net/isc-project}. 
 The sample has been simulated with the \enzo \ code \citep{ENZO_2014} using the the piecewise parabolic method hydro solver \citep[][]{1984JCoPh..54..174C}.
 We applied the WMAP7 $\Lambda$CDM cosmology \citep{2011ApJS..192...18K} in our simulations: $\Omega_0 = 1.0$, $\Omega_{\mathrm{B}} = 0.0445$, $\Omega_{\mathrm{DM}} = 0.2265$, $\Omega_{\Lambda} = 0.728$, $h = 0.702$, $\sigma_8 = 0.8$ and a primordial index of $n = 0.961$. 
 Each cluster was extracted from an initial cosmological volume, sampled with $400^3$ cells and $400^3$ dark matter particles, of the size $\approx (63 \ \Mpc)^3$ (co-moving). The central volume $\approx (6.27 \ \Mpc)^3$ around each cluster has been refined further for a final resolution of $\dd x \approx 20 \ \kpc$. The adaptive mesh refinement (AMR) method used in our simulations is the same as described in Sec. 2 of \citet[][]{2017MNRAS.464...210V}. \\
 All simulations started at a redshift of $z = 30$ and  about $\sim 190-250$ data dumps from each simulation, $\sim 160-220$ between redshifts $z = 1$ and $z = 0$, were saved for further analysis. Our simulations are non-radiative and do not include any magnetic fields nor non-gravitational heating, except an imposed temperature floor of $T = 3 \cdot 10^4 \ \K$ to mimic re-ionization at moderate redshifts, e.g. $4 \le z \le 7$. 
 \subsection{\CRaTer}\label{ssec:tracer}
 We use our Lagrangian tracer code  \textit{C}osmic-\textit{Ra}y \textit{T}rac\textit{er} (\CRaTer) \citep[which has already been applied in various works:][]{2017MNRAS.464.4448W,2016MNRAS.459...70V,2016Galax...4...60V,2016Galax...4...71W} to follow the clumpy accretion of gas in post-processing. We use a \textit{Cloud-in-Cell}-method to interpolate the velocity, gas density, temperature, enstrophy and various source terms computed on the \enzo-grid to the tracer's position. The tracers are advected linearly in time. \\
 Following the mass distribution of the \enzo-simulations, the tracers were injected within a volume of $320^3$ cells on the finest grid of the \enzo-simulation at $z = 1$. Using the same mass threshold we injected additional tracers according to the distribution of the mass entering the simulation box during run time. At $z = 0$ each cluster is consequently populated by $\sim 10^6-10^7$ tracers with a mass resolution of $m_{\mathrm{tracer}} \approx 3 \cdot 10^6 \ \Msun$. We choose this mass resolution as it is high enough to resolve structures accurately while the corresponding number of tracers can be still handled computationally.\\
 \section{Results on IT90\_3}\label{sec:results}
\subsection{Cluster properties}\label{ssec:general}
\begin{figure*}
 \includegraphics[width = \textwidth]{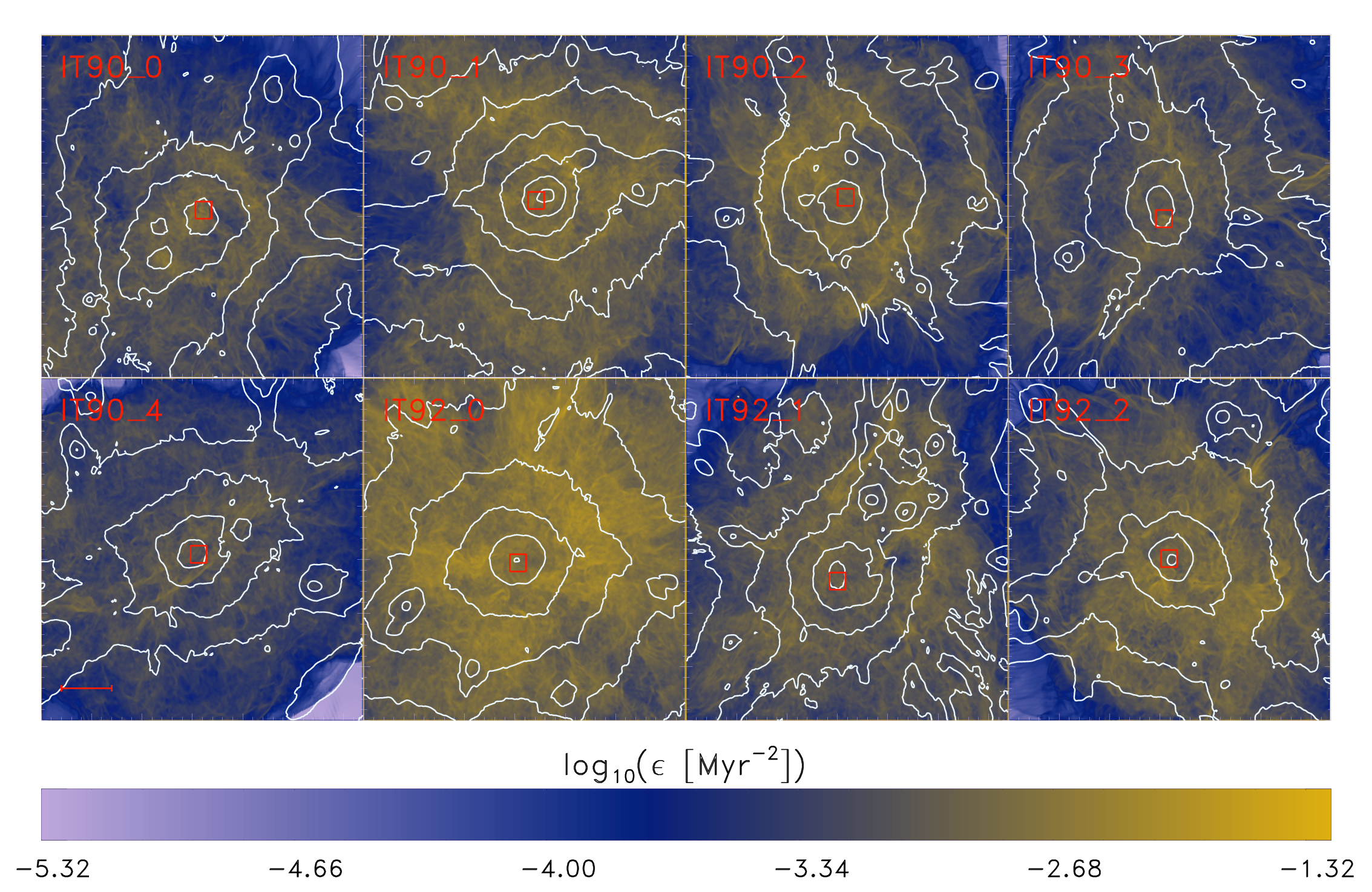}
 \caption{Projected enstrophy overlayed with the density contours of the different clusters in the highes AMR region of size $(6.27 \ \Mpc)^3$ at $z = 0$. The red square in each plot displays the $(320 \ \kpc)^3$ volume surrounding the peak of enstrophy. The red bar show the length of $1 \ \Mpc$. (A coloured version is available in the online article.)}
 \label{fig:ens_maps_all}
\end{figure*}

\begin{figure*}
 \includegraphics[width = 0.76\textwidth]{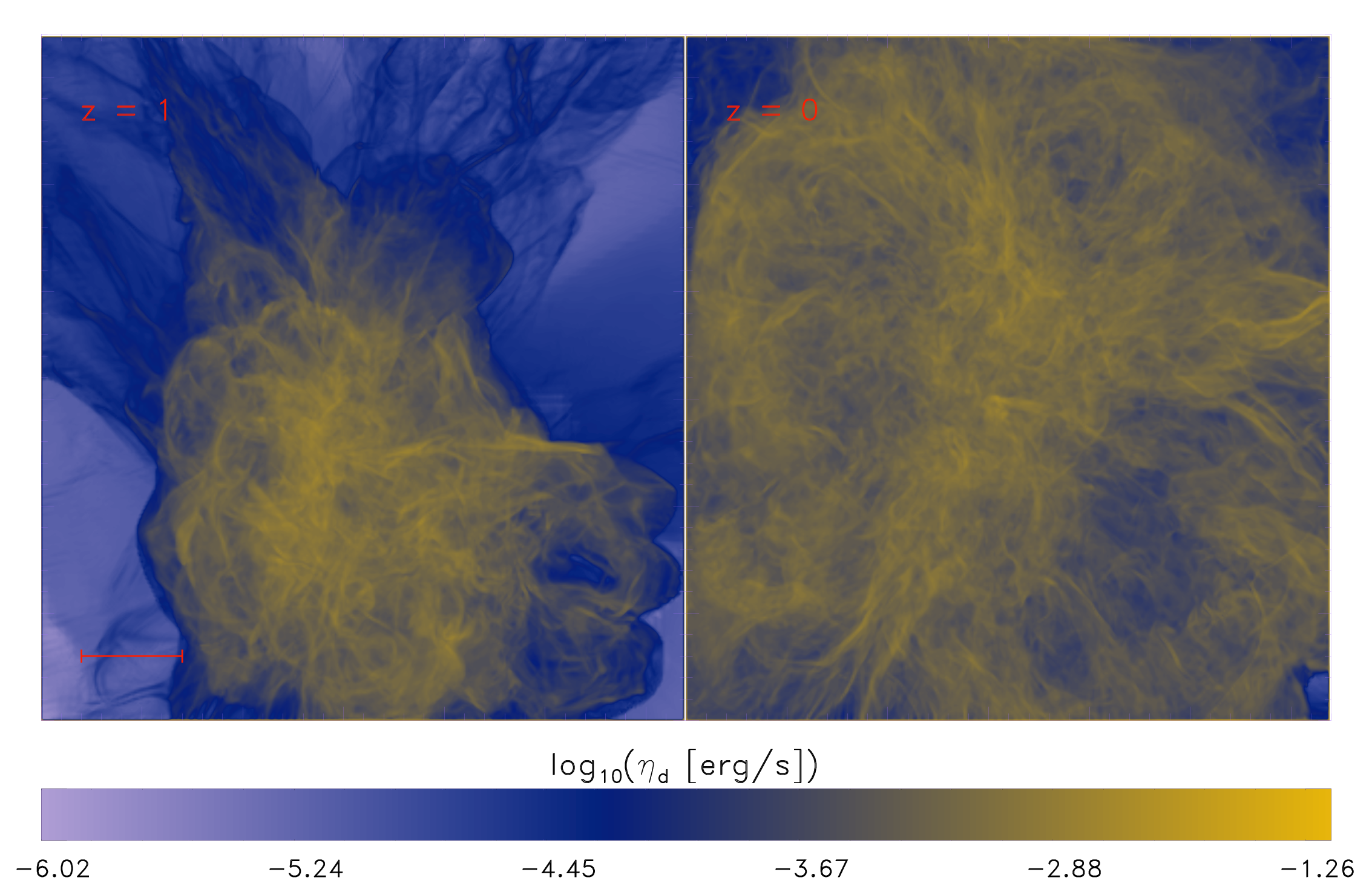}
 \caption{Evolution of the projected enstrophy in the highes AMR region of size $(6.27 \ \Mpc)^3$ of cluster IT90\_3 at $z = 1$ (left) and $z = 0$ (right). The red bar show the length of $1 \ \Mpc$. (A coloured version is available in the online article.)}
 \label{fig:it903ensevo}
\end{figure*}

 At a redshift of $z = 0$ our eight galaxy clusters cover a mass range of  $\Mth = 0.5-3.3 \cdot 10^{14} \ \Msun$ (total mass) and a  temperature range of $T_{200} \approx 5.1 - 19.3 \cdot 10^6 \ \K$, which corresponds to a sound speed range of $266 - 516 \ \km / \sek$. The dynamical and numerical properties of our clusters are summarized in Tab. \ref{tab:cluster}, and a closer look at the dynamical histories and X-ray properties of each individual cluster is given in the appendix \ref{app:dce}.  The classification of each system based on the presence of a major merger has been estimated based on the analysis of the mass accretion history of each system.\\
 The projected enstrophy overlayed with density contours at a redshift of $z = 0$ is shown for all eight clusters in Fig. \ref{fig:ens_maps_all}. The red squares mark the $\sim (320 \ \kpc)^3$ volume centred around the peak of enstrophy, which has been chosen in three dimensions and is therefore not clearly visible in the projected maps. For our tracers analysis we will focus on the tracers that are located in this region at a redshift of $z = 0$.  \\
 Following the methods described in Sec. \ref{ssec:tracer}, we advected tracer particles in post-processing for each cluster between redshifts $z = 1$ and $z = 0$. Most of our discussion will revolve around the merging cluster IT90\_3, which has been already studied in great detail in \citet{2017MNRAS.464...210V}. We will point out differences and similarities with the other ISC clusters where it is most instructive. \\
In Fig. \ref{fig:it903ensevo}, we show the projected enstrophy of cluster IT90\_3 at redshifts $z = 1$ and $z = 0$. At $z = 1$ the enstrophy already spans a range of $10^{-6}-10^{-1}$  Myr$^{-2}$ and fine turbulent structures are visible. Even at the earliest output from our simulation $z = 30$, enstrophy is already at the level of about 1 percent of what it is at $z = 1$ or $z = 0$. \\
\begin{table*}
  \begin{tabular}{l|c|c|c|c|c|c|c|c}
   ID & $M_{\mathrm{200}} \ [10^{14}\cdot \Msun]$ & $r_{\mathrm{200}} \ [\kpc]$ & $T_{\mathrm{200}} \ [10^6\cdot\mathrm{K}]$ & $c_s \ [\km/\sek]$ & major merger & $N_{\mathrm{s}} (z = 30)$ & $N_{\mathrm{s}} (z = 1)$ &  $N_{\mathrm{p}}(z = 0)$ \\
   \hline
   IT90\_0 & $0.77$ &  $881.37$    & $6.88$  & $308$  & no  & $187$ & $156$ & $2.80 \cdot 10^6$ \\
   IT90\_1 & $2.45$ &  $1292.68$   & $10.55$ & $381$  & yes & $194$ & $164$ & $7.75 \cdot 10^6$ \\
   IT90\_2 & $1.10$ &  $998.89$    & $8.29$  & $338$  & no  & $196$ & $166$ & $5.05 \cdot 10^6$ \\
   IT90\_3 & $0.72$ &  $861.78$    & $6.26$  & $293$  & yes & $193$ & $163$ & $4.90 \cdot 10^6$ \\
   IT90\_4 & $0.54$ &  $783.44$    & $5.13$  & $266$  & no  & $197$ & $167$ & $4.07 \cdot 10^6$ \\
   IT92\_0 & $3.32$ &  $1429.78$   & $19.37$ & $516$  & yes & $244$ & $209$ & $8.26 \cdot 10^6$ \\
   IT92\_1 & $1.00$ &  $959.71$    & $7.13$  & $313$  & no  & $227$ & $194$ & $4.82 \cdot 10^6$ \\
   IT92\_2 & $1.17$ &  $1018.47$   & $8.74$  & $347$  & no  & $241$ & $206$ & $4.73 \cdot 10^6$ 
  \end{tabular}
  \caption{Main characteristics of our eight simulated clusters at z = 0: cluster ID, $M_{200}$, $r_{200}$, $T_{200}$, sound speed, dynamical state of the cluster, number of snapshots available between $z = 30$ and $z = 0$ $N_{\mathrm{s}} (z = 30)$, number of snapshots available between $z = 1$ and $z = 0$ $N_{\mathrm{s}} (z = 1)$ and the final number of tracers $N_{\mathrm{p}}(z = 0)$.}
  \label{tab:cluster}
 \end{table*}

\subsection{Evolution of enstrophy}\label{ssec:it90_3_lagrangian}
 \begin{figure*}
 \includegraphics[width = 0.76\textwidth]{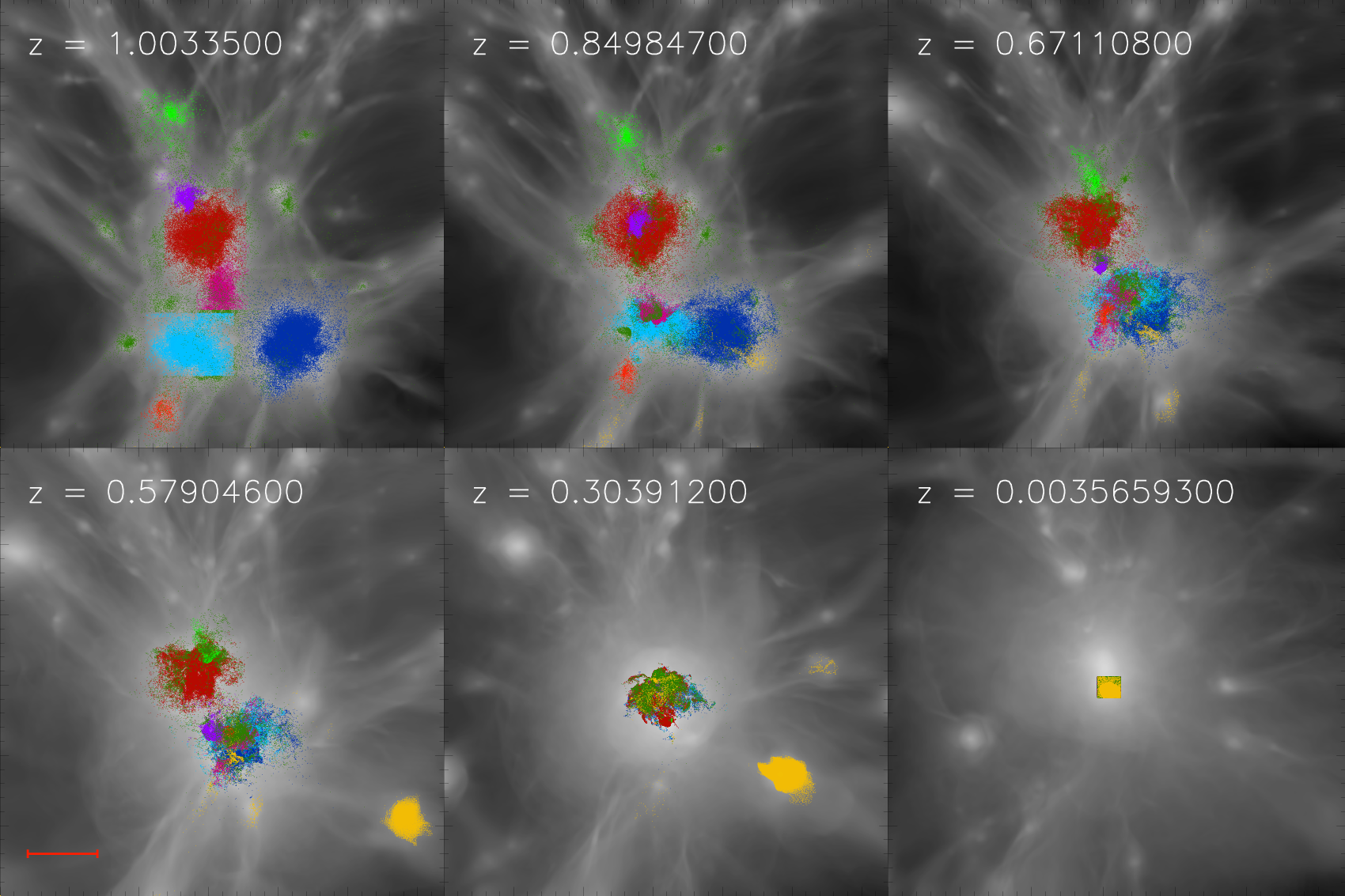}
 \caption{Evolution of the projected gas density overlayed with the tracers position of the different selections in cluster IT90\_3. The tracers have been separated into groups from different subclumps, indicated by the different colours, at $z = 1$. The boxes are of the size $(6.27 \ \Mpc)^3$. The red bar show the length of $1 \ \Mpc$. (See https://dnswttr.github.io/index.html/it903mov.html for a movie. A coloured version is available in the online article.) }
 \label{fig:903_selection_evo}
\end{figure*}
\begin{figure}
 \includegraphics[width = 0.5\textwidth]{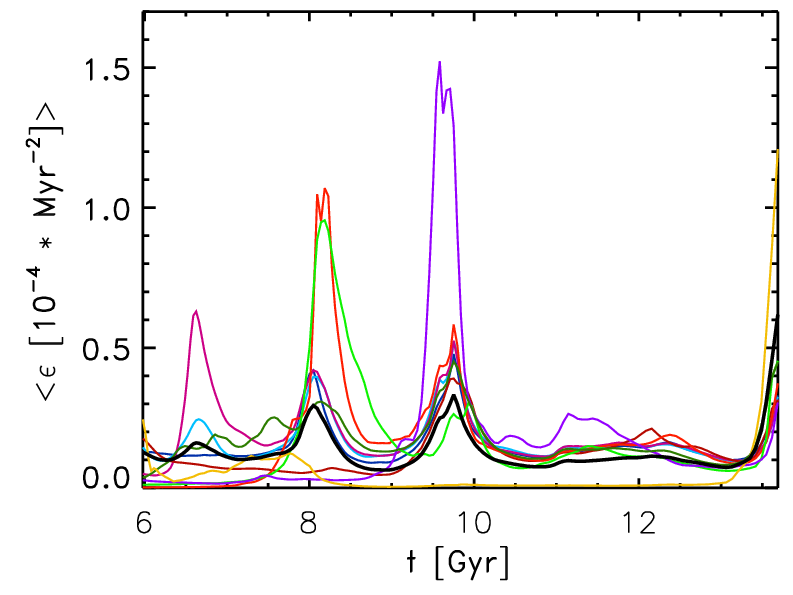}
 \caption{Evolution of the enstrophy, averaged over each tracer family selected in IT90\_3 (colours). The black solid line displays the evolution of enstrophy, averaged over all tracers in IT90\_3. (A coloured version is available in the online article.)}
 \label{fig:903_ens_evo_all_paper}
\end{figure}
\begin{figure*}
 \includegraphics[width = 0.95\textwidth]{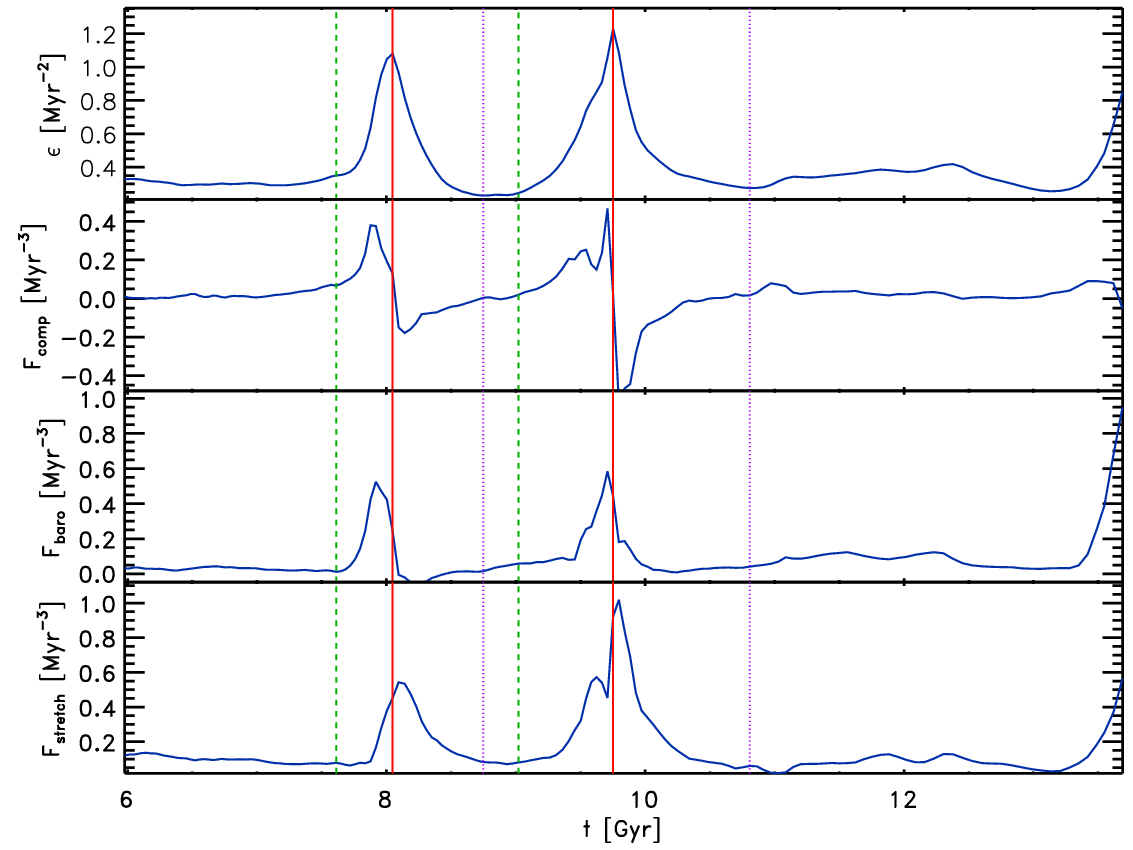}
 \caption{Evolution of $\epsilon$, $\fc$, $\fst$ and $\fb$ of the first family of tracers in IT90\_3 over the last $\sim 7 \ \Gyr$ of the simulation. The red vertical, solid lines mark the local peak of enstrophy, while the green, dashed and purple, dotted lines mark the local minima of enstrophy. (A coloured version is available in the online article.)}
 \label{fig:903_evo_ssub1}
\end{figure*}

\begin{figure*}
 \includegraphics[width = \textwidth]{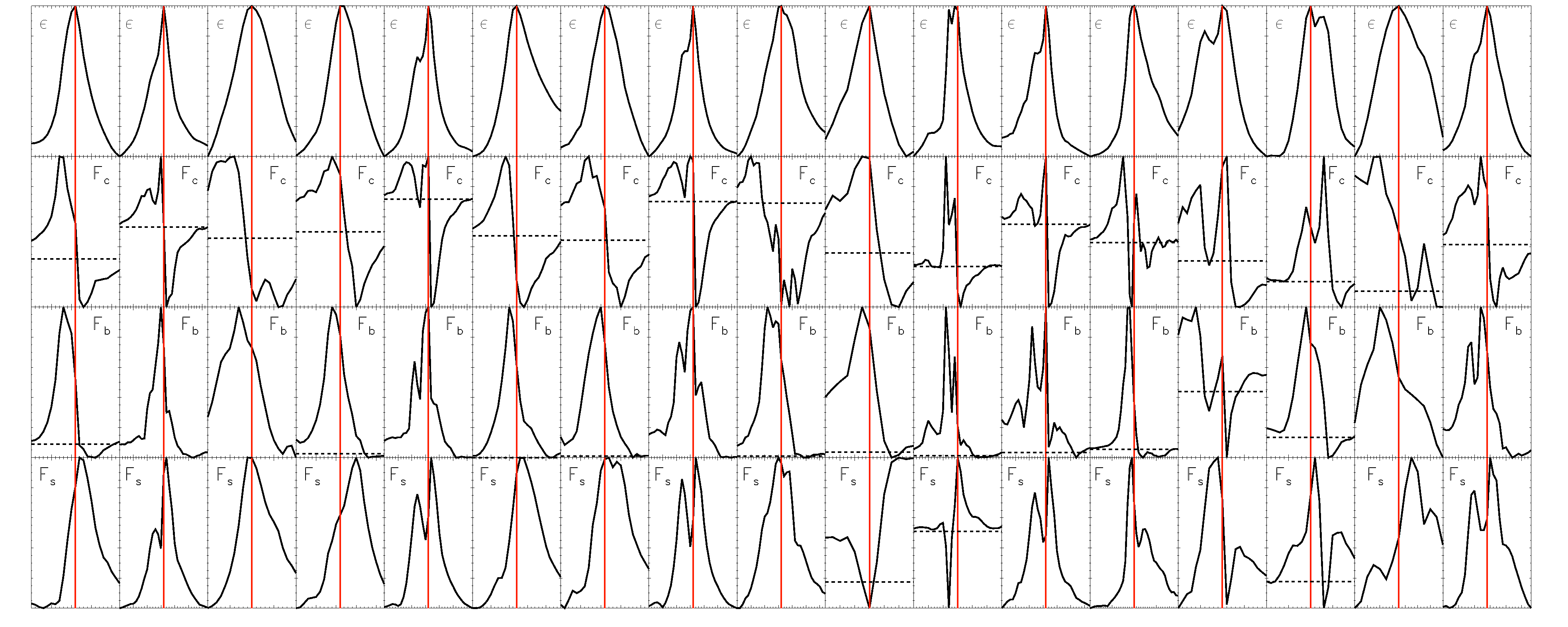} 
 \caption{Summary of all enstrophy ``events'' recorded by \CRaTer \ in IT90\_3. Each column shows a single event recorded by one of the different families. The plots show the evolution of enstrophy (top row), compressive source term (second row), baroclinic source term (third row) and stretching source term (bottom row) around the peaks of enstrophy. The amplitudes (y-axis) of each quantity have been normalized to unity and the time range (width of x-axis) around each each has been normalized to the evolutionary time at the peak of enstrophy. The red line marks the time of the local peak of enstrophy. The black dashed horizontal lines shows the zero level. (A coloured version is available in the online article.)}
 \label{fig:903_enst_vs_F_16_4}
\end{figure*}

\begin{figure*}
 \includegraphics[width = \textwidth]{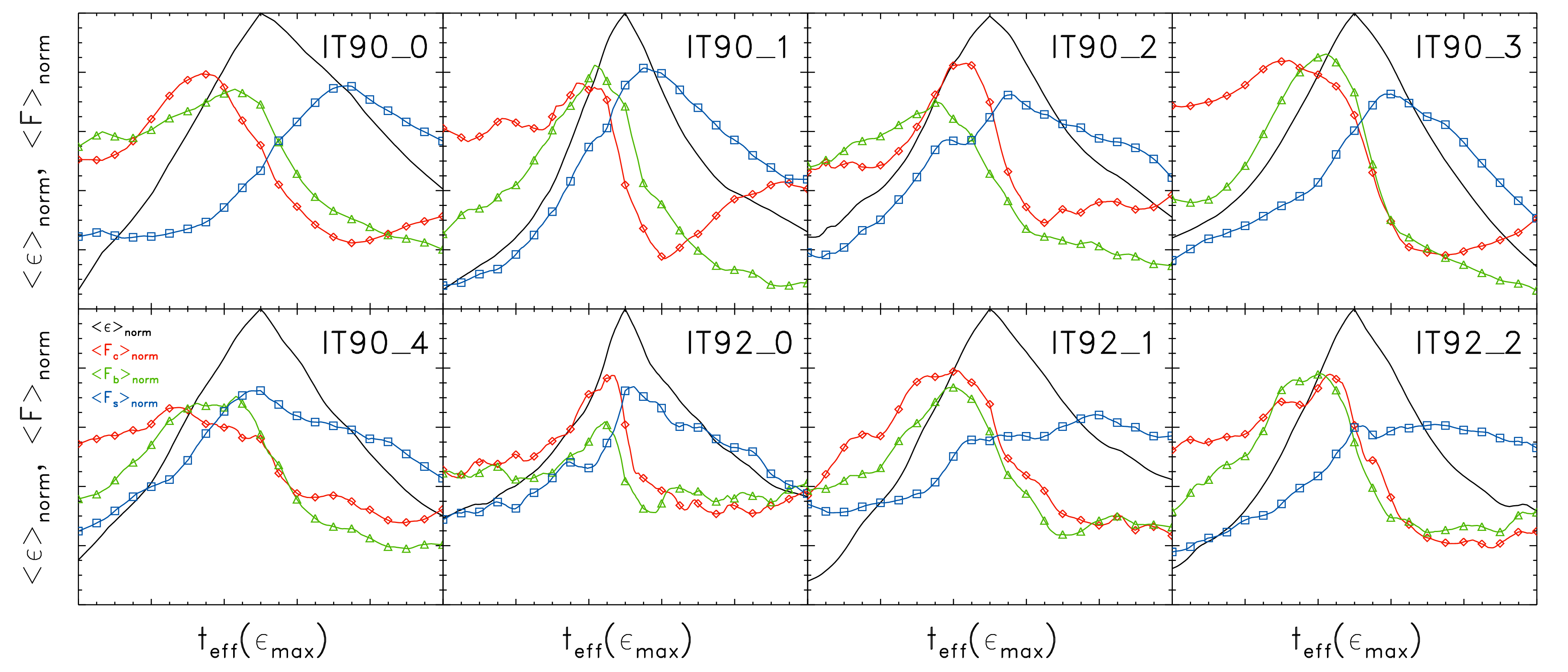}
 \caption{Evolution of the means of enstrophy (black), compressive source term (red, diamonds), baroclinic source term (green, triangles) and stretching source term (blue, squares) normalized to an unit amplitude of one and a unit time equivalent to one evolutionary time measured at the peak of enstrophy, e.g. $t_{\mathrm{eff}}(\epsilon_{\max})$ equals the evolutionary time when the enstrophy is at its maximum. Each panel shows the averages of all events recorded by the tracers in one cluster. It is observed that the compressive and baroclinic source terms are always the strongest before the peak of enstrophy, while the stretching term shows its maximum after the peak of enstrophy. (A coloured version is available in the online article.)}
 \label{fig:IT9_mean_all_in_one_enst_source}
\end{figure*}

          
In order to investigate the source of enstrophy, we selected all tracers in the $\sim(320 \ \kpc)^3$ region centred around the peak of enstrophy at $z = 0$ (see the yellow box in the last panel in Fig.~\ref{fig:903_selection_evo} and red squares in Fig.~\ref{fig:ens_maps_all}). Then we followed the tracer positions back to their positions at $z = 1$.  At that point, most of these tracers are located inside of gas clumps or are entering the high resolution box inside of gas clumps at a later time. Only few tracers cannot be associated with any  gas clump ($\approx 1-10 \%$). At $z = 1$ we further divided the tracers into different families depending on their position (see the different colours in the first panel in Fig. \ref{fig:903_selection_evo}). We selected eight different families of tracers in each cluster, each associated with a gas clump and containing $\sim 10^3-10^4$ tracers plus one additional family that contains all tracers that were injected at the boundaries by mass inflow into the refined volume after $z = 1$. \\
This procedure mostly selects the gas component at $z = 0$ associated with the densest gas substructures in the ICM, which are mostly associated with single self-gravitating gas clumps\footnote{All the clumps have $m_{\mathrm{clump}} > m_{\mathrm{jeans}}$.} that are already formed at $z = 1$ (see Fig. \ref{fig:903_selection_evo}). The total gas masses of these clumps are typically a few $\sim 10^{12} \ \Msun$, corresponding to total masses (gas and dark matter) of a few $\sim 10^{13} \ \Msun$ before ram pressure stripping and tidal interactions detach their gas and dark matter components \citep[e.g.][]{2004MNRAS.350.1397T}. \\
In Fig.~\ref{fig:903_selection_evo}, we show the advection of the selected tracers across cluster IT90\_3. The enstrophy averaged over each individual family of tracers and over all tracers as a function of time is plotted in Fig.~\ref{fig:903_ens_evo_all_paper}. The black line shows the results for all tracers, while the colours correspond to the selection from Fig.~\ref{fig:903_selection_evo}. The mean ensemble enstrophy peaks three times: around $t \approx 6.7, \ 8.1$ and $9.8 \ \Gyr$. The times of the first two events correspond to two minor mergers between sub-clumps, while the time of the third event corresponds to the major merger observed in the IT90\_3 cluster. Using the tracers we can cleanly isolate the different events. We see that peaks of enstrophy (shown in Fig. \ref{fig:903_ens_evo_all_paper}) always occur when two or more tracer families are colliding. As the tracers are following the gas, these events are connected to the merging of clumps. \\
In the following, we will focus on the evolution of four tracer families whose collected enstrophy sharply increases at $t \approx 8.1 \ \Gyr$\footnote{We notice that at the same time the enstrophy of two other families is increased, yet those two families are in a different region at this time and therefore they are related to a different, roughly simultaneous event.}. At  $t \approx 6 \ \Gyr$ $(z = 1)$ the four families are spatially separated. We show the evolution of the group enstrophies in Fig. \ref{fig:903_c5132_corr_a}. The enstrophy of each group always peaks around the time of merging. After the four clumps have merged, they all show the same evolution in enstrophy. The enstrophy peak at $t \approx 9.8 \ \Gyr$ happens during another merger involving these now combined clumps \\ 
 The thermal entropy\footnote{Represented as $S = c\frac{T}{\rho^{2/3}}$, where $c$ is a constant.} (see Fig. \ref{fig:903_c5132_corr_c}) increases significantly when the enstrophy peaks, indicating dissipation either by shocks or by numerical dissipation of turbulence\footnote{For Kolmogorov turbulence it is easy to show from the Navier-Stokes equations that the local turbulent energy dissipation rate scales as $\epsilon^{3/2}$ (see Eq. \ref{eq:etad}) below} itself. The apparent correlation between enstrophy and entropy then suggests that the dissipation of turbulent energy is the dominant mechanism for gas heating here. The evolution of Mach numbers shows strong Mach numbers at the jumps of entropy and enstrophy. 
This supports the idea that these events happen during the occurrence of shocks. \\
 We now examine the individual source terms of the enstrophy (see Eq. \ref{eq:enst_lagrange_deriv}) for the tracer family displayed in dark blue (see Fig. \ref{fig:903_evo_ssub1}). The enstrophy (top row) shows two maxima at $t \approx 8.1 \ \Gyr$ and at $t \approx 9.8 \ \Gyr$, marked by the vertical red lines. The green and purple lines mark the local minima before and after the peak of enstrophy. The compressive and baroclinic source terms (second and third row in Fig. \ref{fig:903_evo_ssub1}) are always the strongest before the enstrophy reaches its maximum. On the other hand, the stretching source term (fourth row in Fig. \ref{fig:903_evo_ssub1}) peaks after the maximum enstrophy. The other selections of tracers in cluster IT90\_3 record the same sequence of events when enstrophy is enhanced (see Fig. \ref{fig:903_enst_vs_F_16_4} for all recorded events). The enstrophy and source terms are normalized to a unit time and unit amplitude. We note that the double peaks in some enstrophy lines (e.g., in the 14th column of Fig. \ref{fig:903_enst_vs_F_16_4}) are numerical artefacts caused by limited time resolution in the ENZO data. \\
\begin{figure*}
 \includegraphics[width = \textwidth]{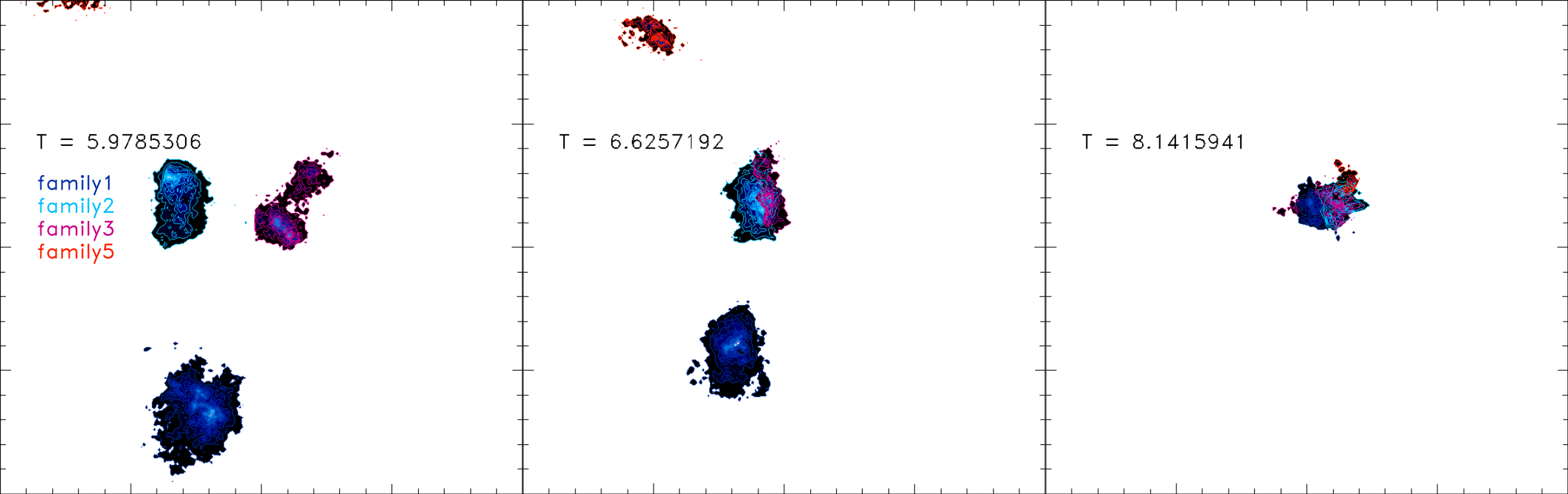}
 \caption{Spatial evolution of four tracer families across cluster IT90\_3. The enstrophy is amplified at the timesteps displayed here due to the merging of the clumps. The displayed regions are of the size $(400 \ \Mpc)^2$.  (A coloured version is available in the online article.)}
 \label{fig:903_approach}
\end{figure*}
\begin{figure*}
 \subfigure[]{\includegraphics[width = 0.49\textwidth]{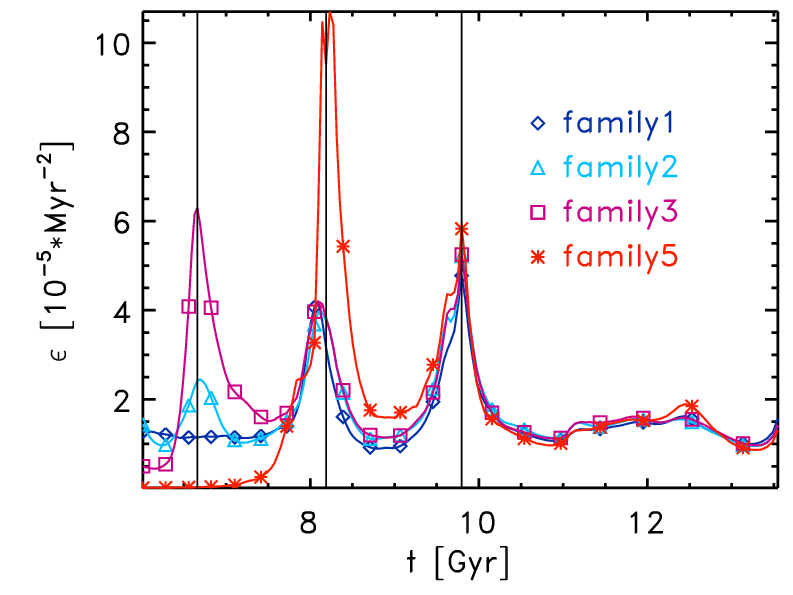}\label{fig:903_c5132_corr_a}} 
 \subfigure[]{\includegraphics[width = 0.49\textwidth]{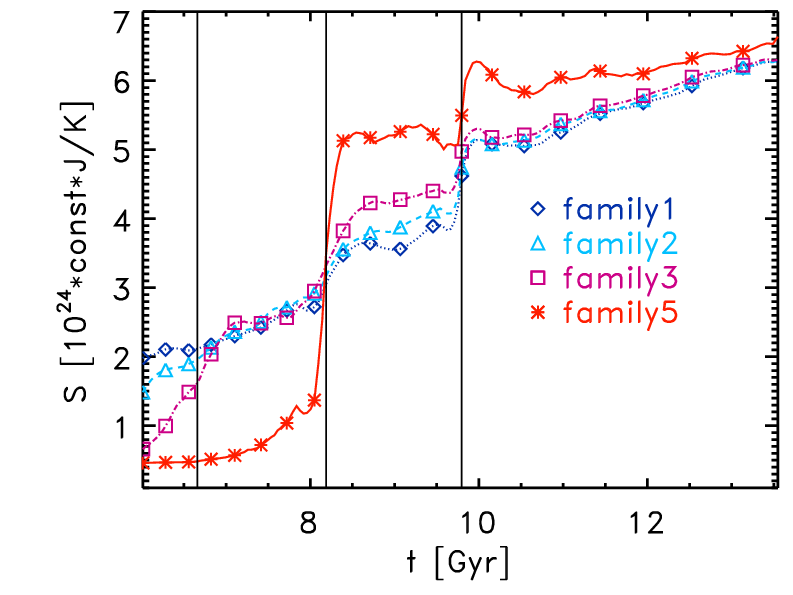}\label{fig:903_c5132_corr_c}} 
 \caption{Evolution of enstrophy in panel (a) and entropy in panel (b) recorded by the four tracer families selected in cluster IT90\_3 and that are shown in Fig. \ref{fig:903_approach}. The black vertical lines mark the timesteps of local maximum enstrophy. (A coloured version is available in the online article.)}
 \label{fig:903_c5132_corr}
\end{figure*}
 In Fig. \ref{fig:IT9_mean_all_in_one_enst_source} we show the mean values of enstrophy and sources terms for IT90\_3 and the other clusters. The same analysis on the other seven clusters gives consistent results: on average the compressive and baroclinic motions are the strongest before the peak of enstrophy, while the stretching source is the strongest after the peak of enstrophy. However, some special cases are pointed out in the following: \\
 IT90\_0 shows the biggest delay between the maximum of compressive/baroclinic source terms (red and green lines in Fig. \ref{fig:IT9_mean_all_in_one_enst_source}) and stretching source term (blue line in Fig. \ref{fig:IT9_mean_all_in_one_enst_source}). As it turns out, the other clusters have much more violent and active histories than IT90\_0, which is our most relaxed cluster. The other clusters have been exposed to a higher dynamical activity disturbing their ICMs. Consequent stretching motions occur throughout the lives of those clusters. Accordingly, the smaller time offsets between the source terms and the less distinct peaks in $\fst$ in IT90\_2, IT91\_1 and IT92\_2 compared to IT90\_0 are the consequences of cluster-scale evolutionary events rather than events related to local clumps. \\
 In the case of IT90\_1, the mean baroclinic source term is significantly stronger than the mean compressive source term. Still the compressive source term peaks before baroclinic source term. This distinction from the other clusters points out that the spatial extent of the compressive source term is much more sensitive to specific structures, e.g. shocks, while the baroclinic source can cover a larger volume since it reflects complex consequences of multiple events in the relatively recent history of the cluster.
\subsection{Growth and decay timescales}\label{ssec:it90_3_tturn}
\begin{figure*}
 {\includegraphics[width = \textwidth]{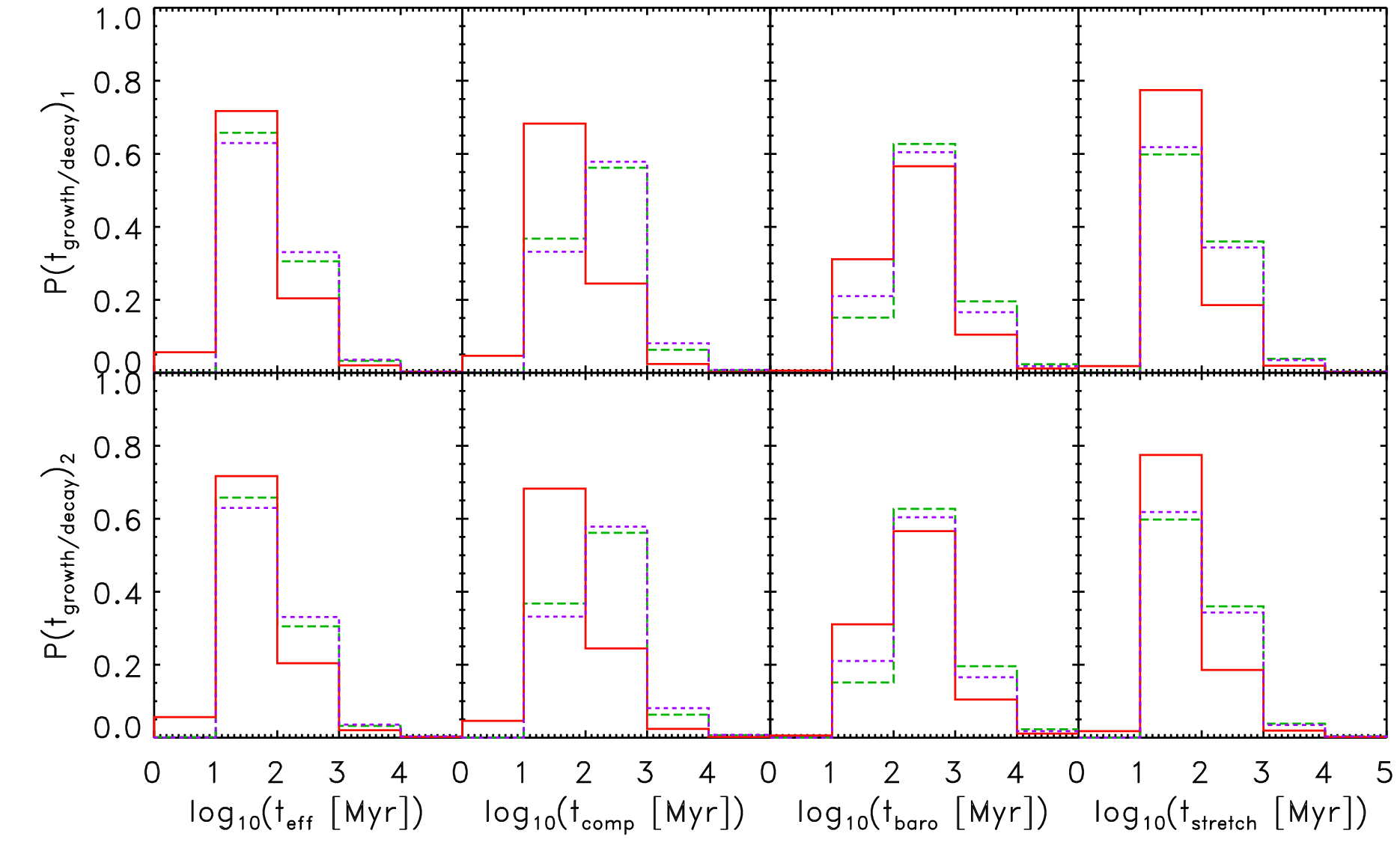}}
 \caption{Distributions of the evolutionary times computed around the times of maximum enstrophy shown in Fig. \ref{fig:903_evo_ssub1}. The top row corresponds to the first peak at $t \approx 8.1 \ \Gyr$ and the bottom row corresponds to the second peak at $t \approx 9.8 \ \Gyr$. The colours and linestyles match the time selections shown in Fig. \protect\ref{fig:903_evo_ssub1}. (A coloured version is available in the online article.)}
 \label{fig:903_tturn_ssub1}
\end{figure*}

\begin{figure}
 \includegraphics[width = 0.5\textwidth]{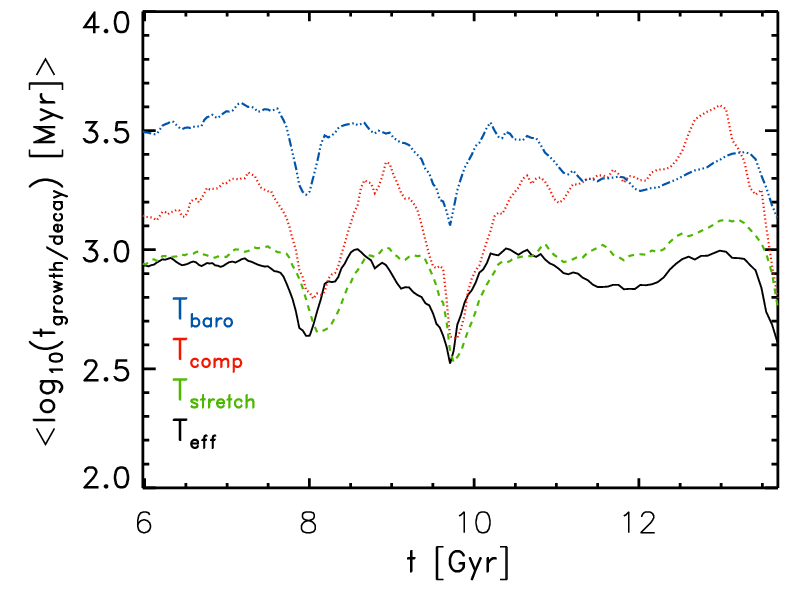}
 \caption{Histories of the effective evolutionary times, see Eq. \ref{eq:Tturn}, (black, solid) and the evolutionary times of the different source terms, see Eq. \ref{eq:Tsource}, recorded by the same selection of tracers shown in Fig. \ref{fig:903_tturn_ssub1}: baroclinic (blue, dash-dotted), compressive (red, dotted) and stretching (green, dashed). (A coloured version is available in the online article.)}
 \label{fig:903_Tturn_color_hist_nobar_subsel_1}
\end{figure}

 Following Eq. \ref{eq:Tturn} and \ref{eq:Tsource}, we estimate the enstrophy growth and decay times\footnote{In the following we will refer to these characteristic growth and decay timescales only as evolutionary times.}, related to the individual source terms and of the effective change of enstrophy, measured by the tracers. We compute the distributions of the different evolutionary times at the six times marked in Fig.~\ref{fig:903_evo_ssub1} (see Fig.~\ref{fig:903_tturn_ssub1}). \\
 At all times the distributions of the effective evolutionary times (left column in Fig. \ref{fig:903_tturn_ssub1}) show the same shape at all times. Most of the tracers recorded an effective evolutionary time in the range of $t_{\mathrm{eff}} \approx 10-100 \ \Myr$ ($\sim 60 \% - 80 \%$ of the tracers) or in the range of $t_{\mathrm{eff}} \approx 10^2-10^3 \ \Myr$ ($\sim 20 \% - 40 \%$ of the tracers). Furthermore we observe that during the events of maximum enstrophy the effective evolutionary times are decreased. \\
 Comparing times for the individual source terms, we notice that $t_{\mathrm{comp}}$ is the most variable. At the peak of enstrophy, the compressive time is mostly in the range of $t_{\mathrm{comp}} \approx 10-100 \ \Myr$ while it is in the range of $t_{\mathrm{comp}} \approx 10^2-10^3 \ \Myr$ at other times. The  evolutionary times of the other source terms remain mostly in a certain time range. The stretching time, $t_{\mathrm{stretch}}$, falls in the range  $\approx 10-100 \ \Myr$ and the baroclinic time, $t_{\mathrm{baro}}$, falls in the range $\approx 10^2-10^3 \ \Myr$. \\
 The analysis of the average evolutionary times (see Fig. \ref{fig:903_Tturn_color_hist_nobar_subsel_1}) shows that stretching motions are dynamically most important for the evolution of turbulence within the cluster. The compressive motions are mostly subdominant throughout most of the clusters' lifetime. Yet, they become important during the shock-related amplification of turbulence. The baroclinic source term on the other hand only has a small dynamical impact. While baroclinicity is an essential source of enstrophy, it is mostly a minor contributor to the net growth of enstrophy in comparison to compression and especially to stretching (see Fig.~\ref{fig:903_Tturn_color_hist_nobar_subsel_1}. At very late times, in this cluster the baroclinic source term becomes competitive with the compressive source term for a short amount of time (see $ 12 \ \Gyr < t < 13 \ \Gyr$ in Fig. \ref{fig:903_Tturn_color_hist_nobar_subsel_1}), perhaps because baroclinic contributions are more broadly distributed in that era. This is because baroclinic sources are concentrated in shocks, which are relatively weak after the last merger event. On the other hand, the contribution from the baroclinic source term is negligible in dense environments (as it is always smaller than the solenoidal source term). However, in \citet[][]{2017MNRAS.464...210V} we showed that it gets very important for the enstrophy generation in cluster outskirts, where flows following oblique shocks first inject vorticity in the ICM. \\
 The evolutionary times in the other clusters besides IT90\_3 show the same qualitative behaviour. The stretching source term always shows the shortest evolutionary time, in the range of $t_{\mathrm{eff}} \approx 10-100$, while the other source terms show a comparable evolutionary time around the major events that amplify enstrophy and they show a larger evolutionary time otherwise. We also notice that the more relaxed clusters, e.g. IT90\_0, show large evolutionary times of around $t_{\mathrm{eff}} \approx 10^2-10^3 \ \Myr$. \\
 In summary, our analysis shows that the fastest stage of enstrophy evolution of the densest substructures in the ICM is  dominated by compression and shortly followed by stretching of vorticity. The baroclinic generation of vorticity is less important in this density regime, but it produces substantial vorticity at earlier times and across outer accretion shocks. \\
\subsection{Dissipation term and magnetic field amplification}\label{ssec:diss}
\begin{figure*}
 \subfigure[]{\includegraphics[width = 0.49\textwidth]{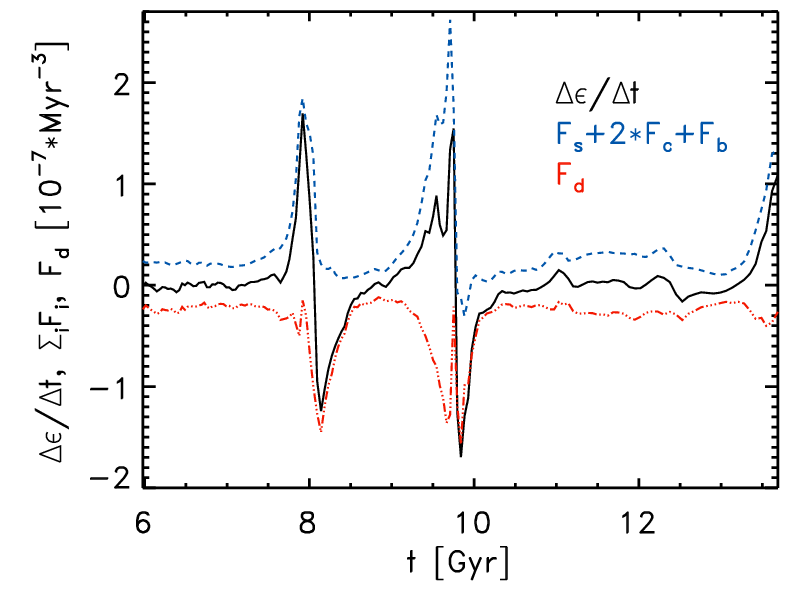}\label{fig:903_dedt_estimate_Fd}}
 \subfigure[]{\includegraphics[width = 0.49\textwidth]{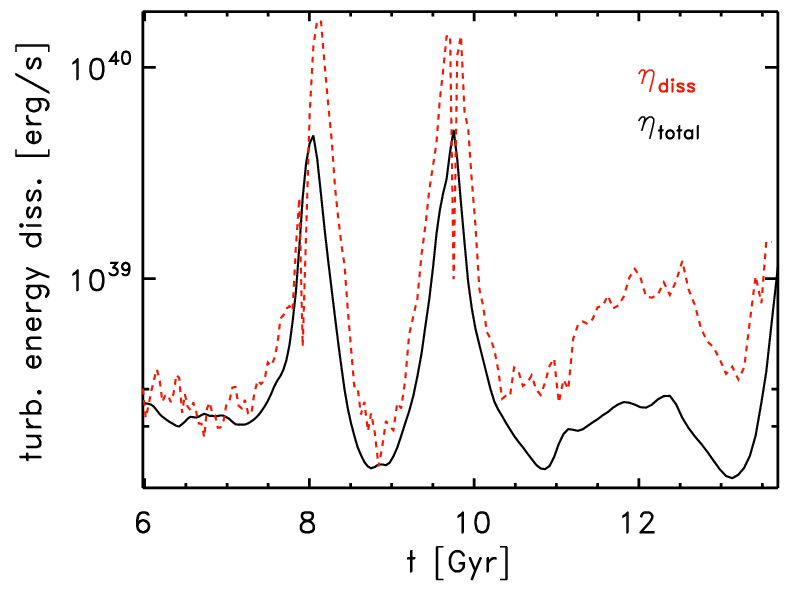}\label{fig:903_diss_eflux}}
 \caption{Panel (a): Evolution of the left (black, solid) and right (blue, dashed) handside of Eq. \ref{eq:enst_lagrange_deriv}. The red line shows the difference of the two, which we associate with viscous dissipation, see label $\fdiss$ (red line, dash-dotted). (b): Comparison of $\eta_{\mathrm{diss}}$ (red, dashed), computed with $\fdiss$ from Eq. \ref{eq:fdiss}, with the enstrophy dissipation rate computed from Eq. \ref{eq:etad} (black, solid). (A coloured version is available in the online article.)}
 \label{fig:903_estimating_dissipation}
\end{figure*}

\begin{figure*}
 \subfigure[]{\includegraphics[width = 0.49\textwidth]{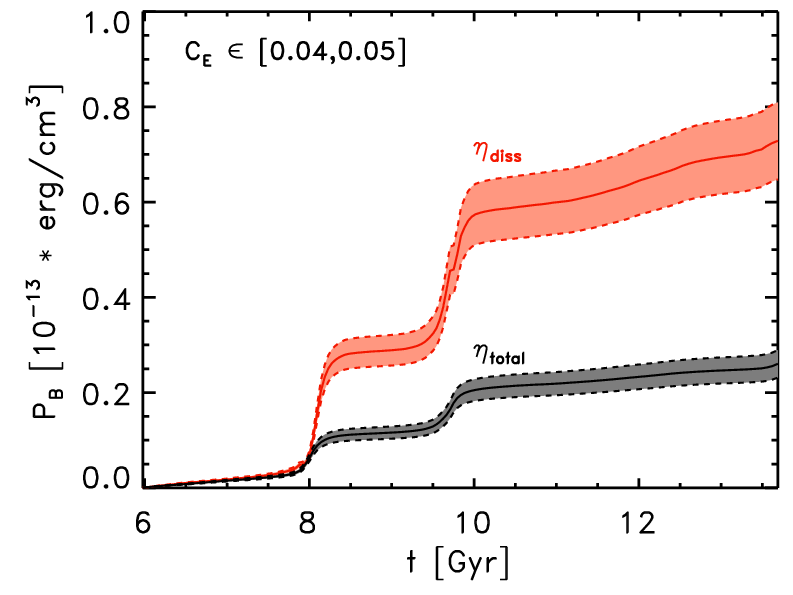}\label{fig:903_etherm_to_emag}}
 \subfigure[]{\includegraphics[width = 0.49\textwidth]{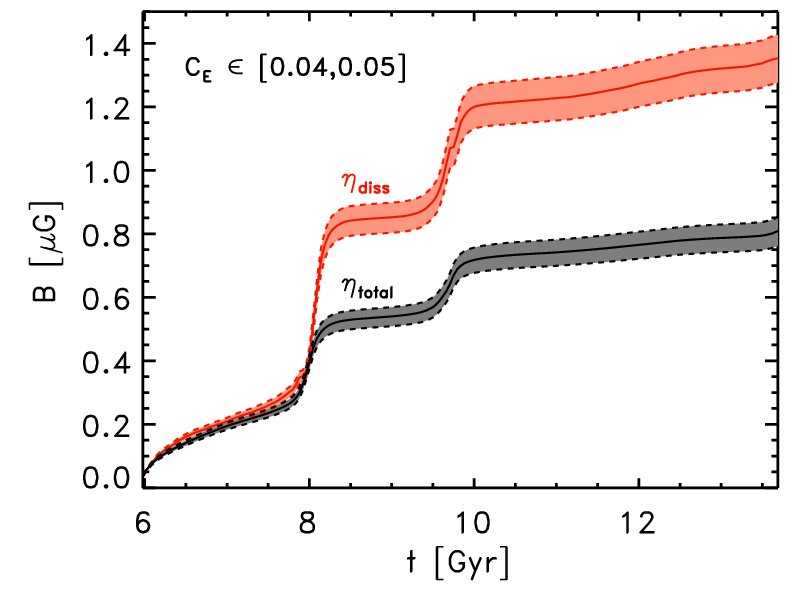}\label{fig:903_etherm_to_bfield}}
 \caption{Panel (a): Integrated magnetic field energy fuelled by the dissipation of turbulence. Panel (b): The corresponding magnetic field strength. The dashed lines give the lower and upper limit obtained with efficiencies in the range of $C_E \in [4\%,5\%]$. (A coloured version is available in the online article.)}
 \label{fig:903_etherm_to}
\end{figure*}

In Eq. \ref{eq:enst_lagrange_deriv}, we neglected the dissipation term since it is not well-defined in an ideal flow. However, the numerical dissipation of turbulent motions is obviously present, as we are not employing sub-grid modelling of turbulence \citep[e.g.][]{2015A&C.....9...49S}. Here we try to empirically constrain its amplitude from the offset between the measured enstrophy change and the summed source terms in Eq. \ref{eq:enst_lagrange_deriv}. In Fig. \ref{fig:903_dedt_estimate_Fd}, we plot the evolution of the right (blue line) and left (black line) hand-side of Eq. \ref{eq:enst_lagrange_deriv} computed with the same tracer family that we have studied in detail in the previous sections and that is displayed in Fig. \ref{fig:903_evo_ssub1}. The two evolutions show a non constant offset, which we plausibly ascribe to the effect of dissipation. Especially at $t \approx 8.2 \ \Gyr$ and $t \approx 10 \ \Gyr$ the difference is not constant showing that turbulence is dissipated.  Hence we compute the dissipation term as:
 \begin{align}
  \fdiss = \frac{\Delta \epsilon}{\Delta t}- \left(2 \cdot F_{\mathrm{comp}} + F_{\mathrm{stretch}} +  F_{\mathrm{baro}}\right). \label{eq:fdiss} .
 \end{align}
 The amount of dissipated enstrophy is the time-integrated absolute value of Eq. \ref{eq:fdiss} $\epsilon_{\mathrm{diss}} = \int \left| \fdiss \right| \dd t $. In the Navier-Stokes formalism (see App. \ref{app:math}) the dissipation rate can be computed without any knowledge on the explicit viscosity. In fact, to a first approximation only a minimum turbulent scale is required (which admittedly depends on the existence of an effective viscosity). In our case, this minimum scale is set by the cell size of our grid cells. Hence, we can compare the results of Eq. \ref{eq:Fdiss} with the net effective dissipation. Following the approach of \citet{2017MNRAS.464...210V}\footnote{See also App. \ref{app:math}.}, we compute this as:
 \begin{align}
  \bar{\eta}_{\mathrm{i}} = 0.014 \cdot \epsilon_i^{\frac{3}{2}} \cdot l_{\mathrm{turb}}^2 \label{eq:etad} ,
 \end{align}
using a length scale\footnote{This is the width of the stencil used to compute the vorticity and which also represents an approximation to the minimum scale for the cascade.} of $l_{\mathrm{turb}} = 2 \cdot \dd x$ and the total amount of enstrophy $\epsilon_i$. In Fig. \ref{fig:903_diss_eflux} we compare the mass-integrated values, $\eta_{\mathrm{i}}$, of both quantities computed using the tracers. We observe that they are in general agreement (see Fig. \ref{fig:903_dedt_estimate_Fd}) and assume that $\fdiss$ in Eq. \ref{eq:fdiss} is a reasonable proxy for the dissipation rate of the turbulent cascade in our simulations. \\
 If the ICM is magnetised and the gas flow is turbulent enough to produce a small-scale dynamo, a fraction of the turbulent energy in the ICM is transferred to the intracluster magnetic fields \citep[for recent reviews see e.g.][]{2006MNRAS.366.1437S,2008Sci...320..909R,2012SSRv..166....1R}. In a predominantly sub-sonic, non stationary and solenoidal turbulence the amplification of ICM magnetic fields can substantially differ from what measured in the simulated ISM \citep[e.g.][]{2016ApJ...825...30P,2017NJPh...19f5003K}. Recently, \citet{2015Natur.523...59M} estimated the efficiency of turbulent energy that is transferred to magnetic fields to be in the range of $C_E \approx [4\%, 5\%]$. Following their approach we compute the evolution of the magnetic energy from the history of turbulent dissipation as (see App. \ref{app:math})
 \begin{align}
  E_B(t) = \frac{B^2}{8 \pi}  =C_E \int^t \rho \eta_{\mathrm{i}}(t') \dd t'.
 \end{align}
 As an example, we estimate the evolution of magnetic energy based on enstrophy evolution by one selection of tracers from IT90\_3 in Fig. \ref{fig:903_etherm_to_emag}. At $t \approx 13.1 \ \Gyr$ the magnetic energy is of the order of $E_B \approx 0.8 - 1.3 \cdot 10^{-13}$  erg cm$^{-3}$ using $\epsilon_{\mathrm{diss}}$ and of $E_B \approx 0.3 - 1.4 \cdot 10^{-13}$ erg cm$^{-3}$ using $\epsilon_{\mathrm{total}}$. This translates into magnetic fields ranging around $B \approx 1.4-1.8 \ \mu \G$ and $B \approx 0.85-1.05 \ \mu \G$, respectively. The above estimates for the magnetic field strengths were computed using the tracers that reside in the cluster core region at $z = 0$. The values estimated in this way are in good agreement with results from observations \citep[e.g.][]{2010A&A...522A.105G}. Both, the magnetic field and the magnetic energy are increased stepwise at $t \approx 8 \ \Gyr$ and $t \approx 10 \ \Gyr$ tracing the evolution of the dissipation term. The timing of these jumps coincide with the times of the merging events in IT90\_3. During the mergers, enstrophy is quickly amplified and is then rapidly dissipated again. Part of this energy will be transferred to the magnetic fields on eddy turnover timescales. The magnetic field growth becomes slower soon after the turbulence subsides. For our estimates on the magnetic field amplification, we neglected magnetic field dissipation that becomes important once the turbulence decays. Therefore, our results are an upper limit. \\
 We observe similar results in the other clusters of our sample. In all clusters, we estimated the mass-integrated values of $\eta_{\mathrm{diss}}$ and $\eta_{\mathrm{total}}$ to be of the same order, which are both in the range of $10^{38}-10^{40}$ erg s$^{-1}$. Application of the above model for transfer of solenoidal turbulent energy into magnetic energy produces magnetic fields, this will produce magnetic fields of the order of a few $\mu \G$. In all cases we observe the episodic jumps in the magnetic field growth. These jumps are always connected to some kind of merging activity. \\
\section{Discussion \& conclusions}\label{sec:summary} 
 We studied the origin and history of enstrophy of the ICM in galaxy clusters formed in Eulerian grid cosmological simulations. We did this using Lagrangian tracer particles that tracked the evolution of the enstrophy in their associated ICM mass along with the associated enstrophy sources and sinks. This way we could analyse the accretion history of turbulence in eight clusters with different formation histories and with different dynamical states at $z = 0$. In the Lagrangian frame enstrophy sources and sinks can be decomposed into compressive, baroclinic, stretching and dissipative terms derived from the compressible Navier-Stokes equations and as defined in Eq. \ref{eq:enst_lagrange_deriv}. We used the Lagrangian tracer code \CRaTer \ to follow these different source terms and the relative timings of cluster formation events at the peaks of enstrophy. \\
 Close examination of  the gas flow properties early in our simulations revealed that small small amount of enstrophy, and therefore turbulence in gas clumps, has already been generated at an early age of the cluster, $z \gg 1$. We have not determined the origin of this very early enstrophy, yet we consider that at least some of it may result from baroclinicity in the cosmologically-based simulation initial conditions. Our analysis showed that additional enstrophy is later generated by baroclinic motions resulting from shocks during the cluster evolution. The generated enstrophy is amplified by compressive and stretching motions. Enstrophy, in association with the turbulence, is dissipated on small scales, just as its turbulent kinetic energy. This turbulent energy contribution contributes substantially to heating of the ICM. Our tracer analysis showed that there is a clear sequence of cluster formation events that lead to strong amplification and decay of enstrophy. During merger events we observed first an increase in the compressive source term, indicating that compression that is mostly connected to shocks is amplifying the enstrophy. Around the same time the baroclinic source term is growing as well, supporting the connection to shocks, and additional enstrophy is generated. Following these two developments the enstrophy reaches its maximum and then starts to decay again. From the previous discussion and results in App. B, we see that the enstrophy dissipation rate increases strongly as the enstrophy increases ($\fdiss \propto \epsilon^{3/2}$), so once $\fb$ and $\fc$, which are the primary solenoidal turbulence drivers, diminish, the dissipation rate overwhelms even a strong $\fst$ source and $\epsilon$ decays along with the solenoidal turbulent energy.\\
 In order to obtain a more quantitative view of the dynamical importance of each source term over time, we computed the effective and individual evolutionary time of the source terms. 	Throughout the whole cluster history, the stretching source term has on average the shortest evolutionary (the fastest enstrophy amplification) time with $t_{\mathrm{stretch}} < 10^3 \ \Myr$ and therefore enstrophy amplification is largely controlled by is controlled by stretching. This seems natural as vortex stretching and energy dissipation are independent of the fluid viscosity, e.g. the dissipative anomaly, in incompressible turbulence. On the other hand, the compressive and baroclinic evolutionary times range between $t_{\mathrm{baro, \ comp}} > 10^3 \ \Myr$ during most of the cluster lifetime making them weak compared to the stretching source term.  They only become competitive, when they are $t_{\mathrm{baro, \ comp}} < 10^3 \ \Myr$ during dynamical events when shocks and other compressions are strong, such as during mergers. This is consistent with our results that the stretching motions are dynamically most important for the evolution of turbulence in galaxy clusters. Yet, baroclinic motions are needed to initially generate turbulence and compressive motions are, once they are acting, a strong booster for enstrophy. The above results are consistent for all clusters that we examined.\\
 The enstrophy dissipation rate peaks when the enstrophy peaks, as already noted. This situation also corresponds to the most rapid amplification of ICM magnetic field and, quite possibly, the peak rate of turbulent acceleration of cosmic rays \citep[see][and references therein]{Brunetti_Jones_2014_CR_in_GC}. In the case of magnetic fields, using magnetic field behaviors from existing MHD turbulence simulations we estimated peak ICM magnetic field strengths $\sim ~\mu$G in our simulated clusters, consistent with estimates from current radio observations \citep[e.g.][]{2010A&A...522A.105G}.\\
 As a final remark, we notice that the study of the internal dynamics of gas substructure is very relevant to model high-resolution X-ray observations of  groups falling onto larger clusters \citep[e.g.][]{2000ApJ...541..542M,2008ApJ...688..208R,2014A&A...570A.119E,2015MNRAS.448.2971I,2016A&A...592A.154D} and their implication to understand plasma processes in these environments. More work is also need to investigate the effects of cooling, feedback \citep[e.g.][]{2009MNRAS.399..497D} and gas viscosity \citep[e.g.][]{2015ApJ...806..104R}, which were not included in this work.
\section*{acknowledgements}
We thankfully acknowledge G. Brunetti and D. Eckert for fruitful scientific conversations.  DW acknowledges support by the Deutsche Forschungsgemeinschaft (DFG) through grants SFB 676 and BR 2026/17. TWJ acknowledges support from the US NSF through grant AST121159.  FV acknowledges personal support from the grant VA 876/3-1 from the DFG, and from the European Union's Horizon 2020 research and innovation programme under the Marie-Sklodowska-Curie grant agreement no.664931. FV and MB also acknowledge partial support from the grant FOR1254 from DFG.  The \enzo-simulations have been carried out in the \textit{ITASCA}-cluster hosted by the University of Minnesota.
 \bibliographystyle{mnras}
 \bibliography{mybib}
\appendix
\section{Cluster dynamics}\label{app:dce}
\begin{figure}
  \includegraphics[width = 0.5\textwidth]{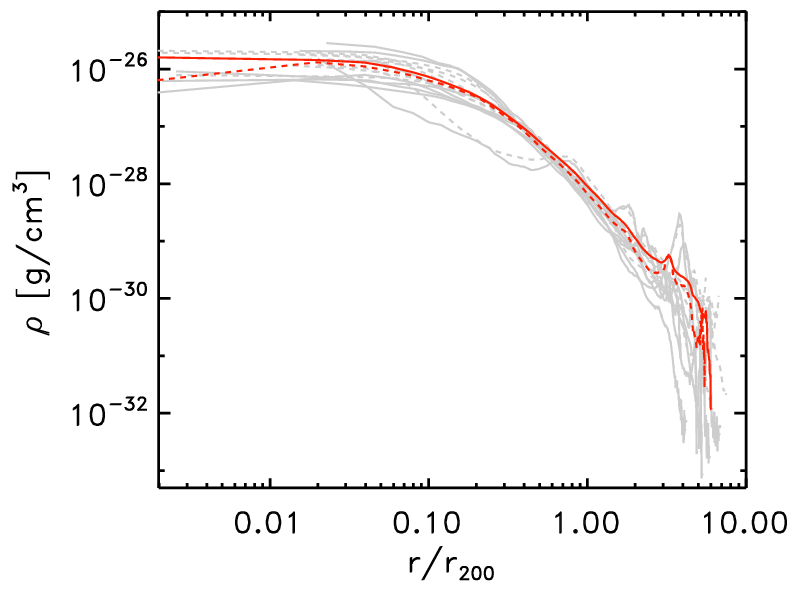}
  \caption{Radial density profiles of the clusters at $z \approx 0$. The solid lines give the results of the Eulerian, unweighted grid average and the dashed lines give the Lagrangian tracer particle-weighted average. The red lines show the average over all clusters. (A coloured version is available in the online article.)}
  \label{fig:all_prof}
 \end{figure}

\begin{figure*}
 \includegraphics[width = \textwidth]{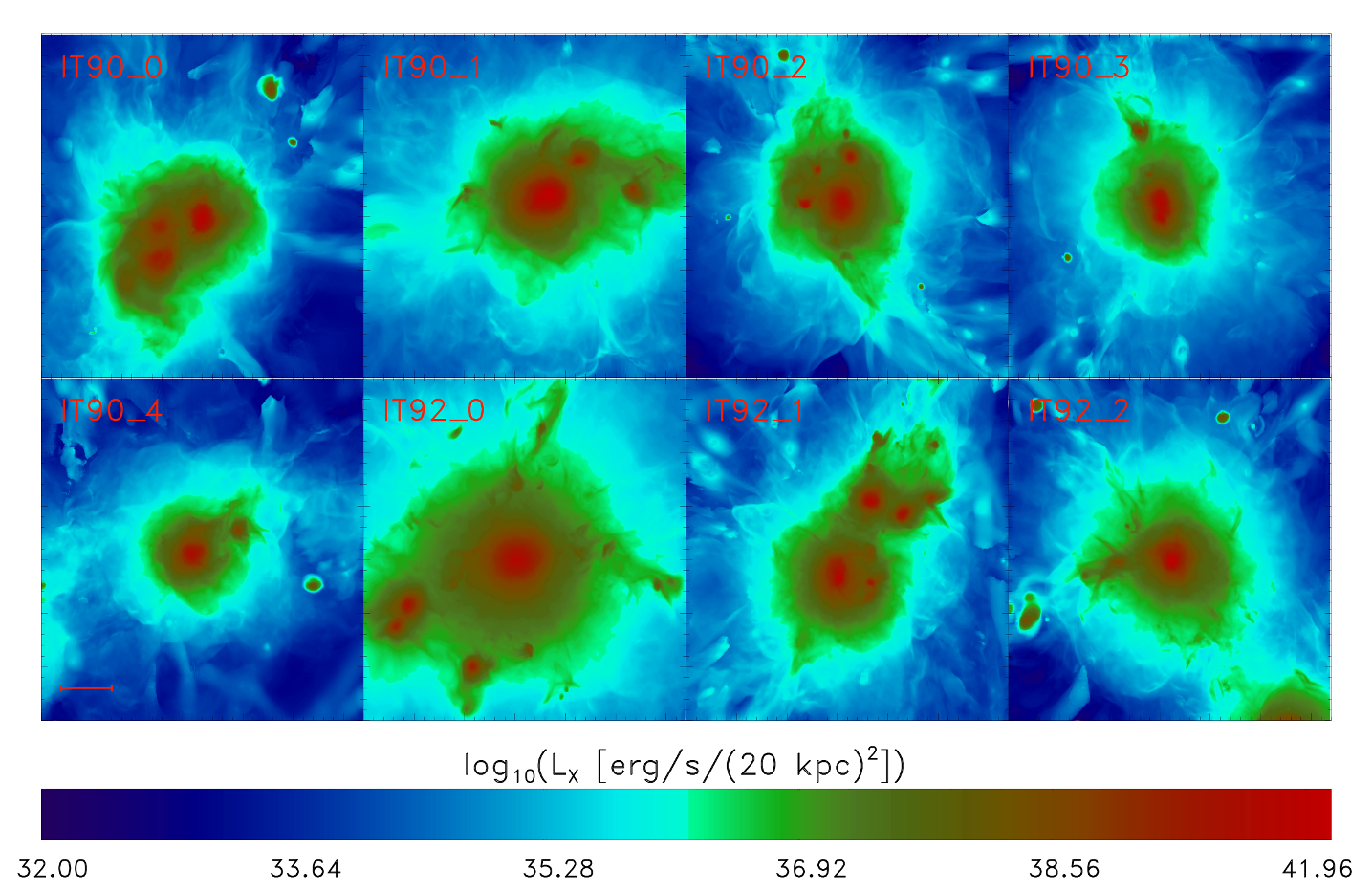}
 \caption{Projected X-ray surface brightness along the line of sight of all the clusters of our sample. Each box is of the size $\approx (6.27 \ \Mpc)^3$ with an resolution of $\dd x \approx 20 \ \kpc$. The red line in the panel of IT90\_4 show the length of $1 \ \Mpc$. The red bar show the length of $1 \ \Mpc$. (A coloured version is available in the online article.)}
 \label{fig:Lx_maps_all}
\end{figure*}

\begin{figure*}
 \includegraphics[width = \textwidth]{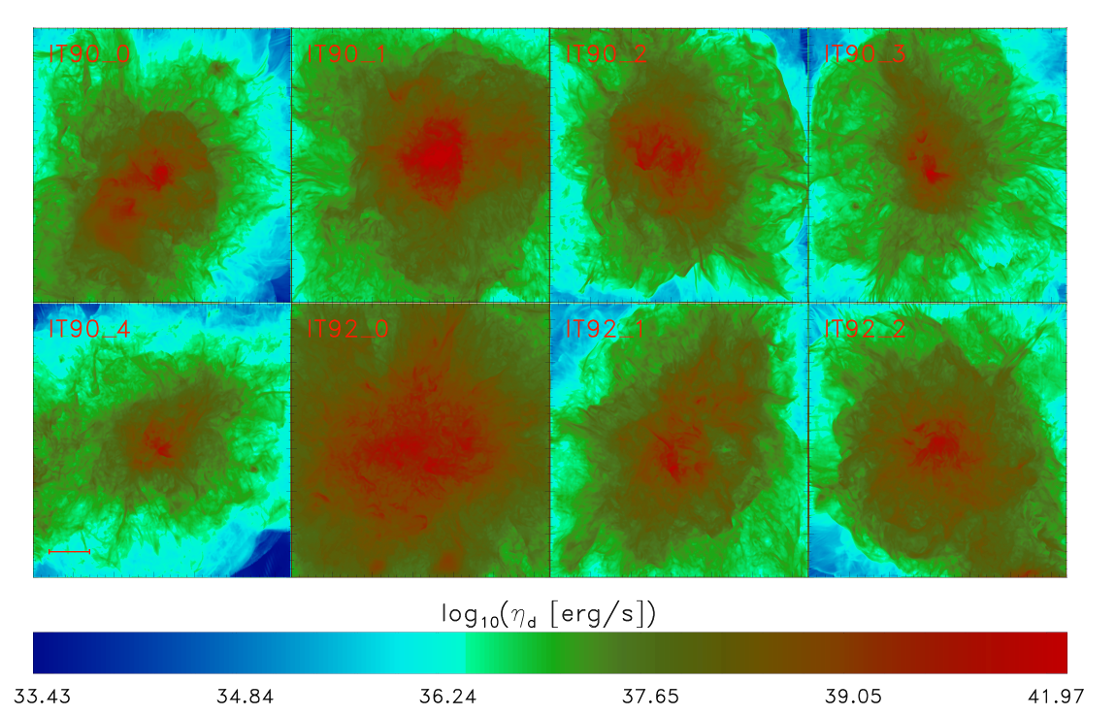}
 \caption{Projected turbulent energy dissipation rate along the line of sight of the clusters contained in our sample. Each box is of the size $\approx (6.27 \ \Mpc)^3$ with an resolution of $\dd x \approx 20 \ \kpc$. The red bar show the length of $1 \ \Mpc$. (A coloured version is available in the online article.)}
 \label{fig:dens_maps_all}
\end{figure*}

\begin{figure}
  \includegraphics[width = 0.5\textwidth]{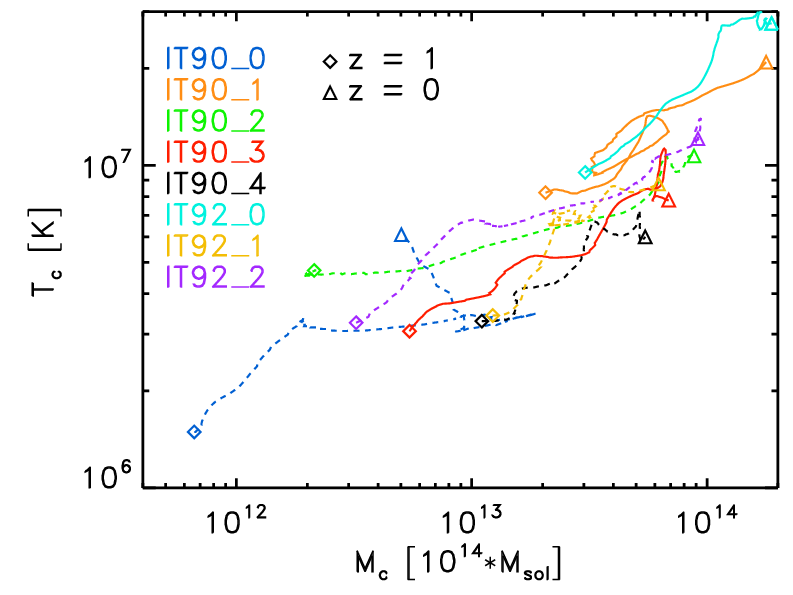}
  \caption{Mass-temperature relation measured in the central $(1.44 \ \Mpc)^3$ of each cluster. The solid lines show the evolution of the major merger clusters and the dashed lines show the evolution of the clusters without a major merger. (A coloured version is available in the online article.)}
  \label{fig:all_hist}
 \end{figure}

 In Fig. \ref{fig:all_prof} we compare the radial density profiles computed with the \enzo \ and \CRaTer \ data at $z \approx 0$. In light grey we show the profiles for each cluster, while the red lines show the average of the profiles over all eight clusters. On the whole the tracers are able to retain the shape of the \enzo \ profile. In Fig. \ref{fig:all_hist} we show the $M-T$ relation of each cluster. The dynamics of the various clusters differ substantially. For example, IT90\_3 hosts a major merger at $t \approx 10.2-10.3 \ \Gyr$ ($z \approx 0.3$), while IT90\_0 stays very relaxed until the end of the simulation. Some clusters, e.g. at at the end of the simulation IT90\_0 or IT92\_1 are on the verge of a major merger, thus accreting a lot of mass, while other clusters, e.g. IT90\_1 or IT92\_0, are only accreting small clumps. \\
  The X-ray surface brightness maps of each cluster are shown in Fig.~\ref{fig:Lx_maps_all}. The cluster centres show a X-ray surface brightness in the range of $L_X = 10^{40} - 10^{42}$ erg s$^{-1} \ (20 \ \kpc)^{-2}$. We show images of the projected turbulent energy dissipation rate (see Eq. \ref{eq:etad}) of the clusters in Fig.~\ref{fig:dens_maps_all} at $z = 0$ ($t \approx 13.72 \ \Gyr$) in a $(6.4 \ \Mpc)^3$ volume. \\

\section{Simple models for turbulent dissipation rates}\label{app:math}

Under the assumption that a turbulent flow with a power law power spectrum can be described as isotropic, solenoidal turbulence, it is possible to express both the kinetic energy dissipation rate, $\eta_d$, and the enstrophy dissipation rate, $F_{diss}$, in forms that do not depend explicitly on the kinematic viscosity, $\nu$. These provide simple and convenient means to estimate the dissipation of turbulence in our simulations, where the viscosity is not well-defined. 

We start from equation \ref{eq:enst_euler_deriv}, which provides an expression for $d\epsilon/dt_{\rm Euler} = \partial\epsilon/\partial t$ obtained from the curl of the compressible Navier-Stokes equation \citep{2015ApJ...810...93P}. The various physical contributions to $d\epsilon/dt_{\rm Euler}$; that is, its source terms, are listed in equations \ref{eq:Fadv} - \ref{eq:Fdiss}. We focus here on the dissipative source term,
\begin{equation}
F_{\rm diss} = \nu\vec\omega\cdot\left(\nabla^2\vec\omega + \nabla\times \vec G\right).\label{appeq:Fdiss}
\end{equation}
Ignoring the strain tensor element $\nabla\times \vec G$, whose predominant role is inside shocks \citep{2015ApJ...810...93P}, we then look for a simple way to estimate
\begin{equation}
F_{diss} \approx \nu\vec\omega\cdot\nabla^2\vec\omega.\label{appeq:Fdiss-approx}
\end{equation}
To obtain estimates of the right hand side of equation (B2) it is useful to utilize the Fourier representation of the turbulent motions. Assuming for simplicity isotropic, Kologorov turbulence in the range [$\ell_1, \ell_o$] it has been shown by many authors \citep[e.g.,][and references therein]{2002physfluid,2011PhRvL.106g5001B} that the turbulent kinetic energy power spectrum can be be expressed in the Fourier domain as
\begin{equation}
E(k) = C_o \eta_d^{2/3} k^{-5/3} = \frac{1}{2}
 v_k^2,\label{appeq:power}
\end{equation}
for $k_o = 2\pi/\ell_o \le k \le k_1 = 2\pi/\ell_1$,
where $\eta_d$ is the turbulent kinetic energy dissipation rate (per unit mass) and $C_o\sim 1.5$ is the so-called Kolmogorov constant. Given that our intent is primarily to establish simple scaling relations, it is not critical whether or not the inner and outer scales in the turbulence are constant across the cluster. The standard expression for viscous kinetic energy dissipation is \citep[e.g.,][]{landau2013fluid}
\begin{equation}
\eta_{\rm d} = 2 \nu ~\sum_{i\ne j} \left(\frac{\partial v_i}{\partial x_j}\right)^2  \label{appeq:l&l}.
\end{equation}
In terms of the Fourier power spectrum, we can then write
\begin{equation}
\eta_{\rm d} =  4\nu \int_{k_o} ^ {k_1}~ k^{2}E(k) dk \label{appeq:l&l2}
\end{equation}
Applying the form for $E(k)$ in equation \ref{appeq:power} we can then obtain a relation for the viscosity, $\nu$ in terms of quantities defining the turbulent power, namely, $C_0$, $\eta_d$ and the range of scales characterizing the  turbulence,
\begin{equation}
\nu \approx \frac{1}{3 C_o} \frac{\eta_{\rm d}^{1/3}}{k_1^{4/3} [1-(\frac{k_o}{k_1})^{4/3}]}.\label{appeq:nuetad}
\end{equation}
Similarly,
\begin{equation}
\fdiss \approx \nu \int_{k_o}^{k_1}  k^2 \omega_k^2 dk \approx   \frac{4}{5} ~\nu~\epsilon k_1^2 ~\frac{1-(\frac{k_o}{k_1})^{10/3}}{1-(\frac{k_o}{k_1})^{4/3}},\label{appeq:fdissnu}
\end{equation}
where $\omega_k =\vec{ k}\times \vec{v_k}$.
Using equation \ref{appeq:nuetad}, equation \ref{appeq:fdissnu}  can be written as
\begin{eqnarray}
\fdiss \approx\frac{1}{5} \left(\frac{4}{3 C_o}\right)^{3/2}~ \epsilon^{3/2}\frac{1-(\frac{\ell_1}{\ell_o})^{10/3}}{[1-(\frac{\ell_1}{\ell_o})^{4/3}]^{5/2}}~~\overrightarrow{\ell_o>> \ell_1}\label{appeq:fdiss-enst}\\\nonumber
 ~\sim  0.17 \epsilon^{3/2}~ [ 1 + (5/2)(\ell_1/\ell_o)^{4/3}].
\end{eqnarray}
In the final expression we assumed $C_o \approx 1.5$. Evidently, the enstrophy dissipation rate is simply $\fdiss\propto \epsilon^{3/2}$, scaled  by a factor that is only moderately sensitive to the ratio of the outer and inner turbulent scales, $\ell_o/\ell_1$. Our empirical estimate for this relation from the IT90$\_$3  cluster gives $\fdiss \approx 0.35 \epsilon^{3/2}$, corresponding to $\ell_o/\ell_1 \sim 31$.
Combining equations \ref{appeq:l&l} and \ref{appeq:nuetad} we can also write the turbulent energy dissipation rate in terms of $\epsilon$ without explicit reference to the viscosity, $\nu$; namely,
\begin{eqnarray}
\eta_d =\left(\frac{4}{3 C_o}\right)^{3/2} \frac{1}{k_1^2}\frac{\epsilon^{3/2}}{1-(\frac{k_o}{k_1})^{4/3}}~~\overrightarrow{\ell_o>> \ell_1} \label{appeq:etad}\\\nonumber
~\sim 0.02 \epsilon^{3/2} \ell_1^2~[1 + (\ell_1/\ell_o)^{4/3}].
\end{eqnarray}
This is also consistent with our empirical estimate for $\eta_d$ in the IT90$\_$3 cluster given in equation \ref{eq:etad}. Note, further, according to equations \ref{appeq:fdiss-enst} and \ref{appeq:etad}  that the ratio $\eta_d/\fdiss\propto \ell_1^2$ with a constant that depends on the ratio of the outer to inner turbulence scales. We note, finally, that even when the turbulence is not truly Kolmogorov, these relations can still provide a useful, if approximate, guide to estimates for the dissipation rates.

\end{document}